\documentclass[sigconf]{acmart}
\usepackage{amsmath}
\usepackage{booktabs}
\usepackage{tabularx}
\usepackage{multirow}
\usepackage{makecell}
\usepackage{natbib}
\usepackage{float}\usepackage{placeins}
\usepackage{booktabs}
\usepackage{tabularx}
\usepackage{array}  
\usepackage{graphicx} 
\usepackage{pifont}
\usepackage{xcolor}
\usepackage {tikz}
\usepackage{transparent}
\usepackage{svg}      
\usepackage[utf8]{inputenc}
\usepackage{geometry}
\usepackage{longtable} 
\usepackage{booktabs}  
\usepackage{array}     
\usepackage{ragged2e}  
\definecolor{BrickRed}{RGB}{203,65,84}  
\definecolor{ForestGreen}{RGB}{34,139,34}

\definecolor{ControlColor}{HTML}{969696}   
\definecolor{GuidanceColor}{HTML}{5C7DB2}  
\definecolor{SuggestionColor}{HTML}{FAC850}
\definecolor{FeedbackColor}{HTML}{2CA02C}  
\definecolor{GenColor}{HTML}{F08080}       

\newcommand{\controlbadge}{\tikz[baseline=-0.5ex]\node[draw=ControlColor, fill=ControlColor!20, rounded corners=2pt, inner sep=2.5pt, font=\tiny\bfseries] {};}
\newcommand{\guidancebadge}{\tikz[baseline=-0.5ex]\node[draw=GuidanceColor, fill=GuidanceColor!100, rounded corners=2pt, inner sep=2.5pt, font=\tiny\bfseries] {};}
\newcommand{\suggestionbadge}{\tikz[baseline=-0.5ex]\node[draw=SuggestionColor, fill=SuggestionColor!100, rounded corners=2pt, inner sep=2.5pt, font=\tiny\bfseries] {};}
\newcommand{\feedbackbadge}{\tikz[baseline=-0.5ex]\node[draw=FeedbackColor, fill=FeedbackColor!100, rounded corners=2pt, inner sep=2.5pt, font=\tiny\bfseries] {};}

\newcommand{\new}[1]{\textcolor{black}{#1}}

\newcommand{\controlbotbadge}{\controlbadge\controlbadge\controlbadge}
\newcommand{\guidancbotbadge}
{\guidancebadge\controlbadge\controlbadge}
\newcommand{\suggestionbotbadge}{\guidancebadge\suggestionbadge\controlbadge}
\newcommand{\feedbackbotbadge}
{\guidancebadge\controlbadge\feedbackbadge}
\newcommand{\genbotbadge}{\guidancebadge\suggestionbadge\feedbackbadge}

\usepackage{listings}
\usepackage{xcolor}

\usepackage{upquote}

\definecolor{cardbg}{HTML}{F8FAFC}
\definecolor{cardframe}{HTML}{D1D5DB}
\definecolor{muted}{HTML}{6B7280}
\definecolor{accentBlue}{HTML}{2563EB}   
\definecolor{accentGreen}{HTML}{16A34A}  
\definecolor{accentAmber}{HTML}{F59E0B}  

\definecolor{kw}{HTML}{1D4ED8}
\definecolor{str}{HTML}{B45309}
\definecolor{com}{HTML}{047857}
\definecolor{emphred}{HTML}{B91C1C}

\lstdefinestyle{paperlist}{
  basicstyle=\ttfamily\scriptsize,
  frame=none,
  breaklines=true, breakatwhitespace=true,
  showstringspaces=false,
  tabsize=2,
  backgroundcolor=\color{cardbg},
  keywordstyle=\color{kw}\bfseries,
  stringstyle=\color{str},
  commentstyle=\color{com}\itshape,
  columns=fullflexible,
  numbers=left,
  numberstyle=\scriptsize\color{muted},
  numbersep=8pt,
  captionpos=b,
  literate={£}{{\textsterling}}1 {’}{\textquotesingle}1,
  moredelim=**[is][\bfseries]{§}{§}, 
  moredelim=**[is][\bfseries\color{emphred}]{@}{@},
}

\lstdefinelanguage{json}{
  basicstyle=\ttfamily\small,
  showstringspaces=false,
  breaklines=true,
  literate=
   *{true}{{{\color{kw}true}}}{4}
    {false}{{{\color{kw}false}}}{5}
    {null}{{{\color{kw}null}}}{4},
  keywordstyle=\color{kw}\bfseries,
  stringstyle=\color{str},
}

\usepackage{listings}
\usepackage{xcolor}

\usepackage[font={bf,sf}, labelfont={bf,sf}]{caption}

\lstnewenvironment{codecardGreen}[1][]
{%
    \lstset{
        style=paperlist,
        backgroundcolor=\color{cardbg},   
        frame=single,                     
        frameround=tttt,                  
        rulecolor=\color{FeedbackColor},  
        framerule=1.5pt,                  
        xleftmargin=2.5em,                
        #1                                
    }
}{}

\lstnewenvironment{codecardBlue}[1][]
{%
    \lstset{
        style=paperlist,
        backgroundcolor=\color{cardbg},
        frame=single,
        frameround=tttt,
        rulecolor=\color{accentBlue},     
        framerule=1.5pt,
        xleftmargin=2.5em,
        #1
    }
}{}

\lstnewenvironment{codecardYellow}[1][]
{%
    \lstset{
        style=paperlist,
        backgroundcolor=\color{cardbg},
        frame=single,
        frameround=tttt,
        rulecolor=\color{SuggestionColor}, 
        framerule=1.5pt,
        xleftmargin=2.5em,
        #1
    }
}{}

\lstnewenvironment{codecardGrey}[1][]
{%
    \lstset{
        style=paperlist,
        backgroundcolor=\color{cardbg},
        frame=single,
        frameround=tttt,
        rulecolor=\color{ControlColor},    
        framerule=1.5pt,
        xleftmargin=2.5em,
        #1
    }
}{}

\AtBeginDocument{%
  }

\setcopyright{none}

\renewcommand{\textsuperscript}[1]{}

\copyrightyear{2026}
\acmYear{2026}
\setcopyright{cc}
\setcctype{by}
\acmConference[CHI '26]{Proceedings of the 2026 CHI Conference on Human Factors in Computing Systems}{April 13--17, 2026}{Barcelona, Spain}
\acmBooktitle{Proceedings of the 2026 CHI Conference on Human Factors in Computing Systems (CHI '26), April 13--17, 2026, Barcelona, Spain}
\acmPrice{}
\acmDOI{10.1145/3772318.3791064}
\acmISBN{979-8-4007-2278-3/2026/04}

\begin{document}

\title{Supporting Effective Goal Setting with LLM-Based Chatbots}


\author{Michel Schimpf}
\email{ms2957@cam.ac.uk}
\authornote{Both authors contributed equally to this research.}
\affiliation{
  \institution{University of Cambridge}
  \city{Cambridge}
  \state{}
  \country{United Kingdom}
}

\author{Sebastian Maier}
\authornotemark[1] 
\affiliation{
  \institution{LMU Munich \& MCML}
  \city{Munich}
  \state{}
  \country{Germany}
}

\affiliation{
  \institution{University of Cambridge}
  \city{Cambridge}
  \country{United Kingdom}
}

\author{Anton Wyrowski}
\affiliation{
  \institution{Technical University of Munich}
  \city{Munich}
  \state{}
  \country{Germany}
}
\affiliation{
  \institution{University of Cambridge}
  \city{Cambridge}
  \country{United Kingdom}
}

\author{Lara Christoforakos}
\affiliation{
  \institution{LMU Munich}
  \city{Munich}
  \state{}
  \country{Germany}
}

\author{Stefan Feuerriegel}
\affiliation{
  \institution{LMU Munich \& MCML}
  \city{Munich}
  \state{}
  \country{Germany}
}

\author{Thomas Bohné}
\affiliation{
  \institution{University of Cambridge}
  \city{Cambridge}
  \country{United Kingdom}
}

\begin{abstract}
Each day, individuals set behavioral goals such as eating healthier, exercising regularly, or increasing productivity. While psychological frameworks (i.e., goal setting and implementation intentions) can be helpful, they often need structured external support, which interactive technologies can provide. We thus explored how large language model (LLM)-based chatbots can apply these frameworks to guide users in setting more effective goals. We conducted a preregistered randomized controlled experiment ($N = 543$) comparing chatbots with different combinations of three design features: guidance, suggestions, and feedback. We evaluated goal quality using subjective and objective measures. We found that, while guidance is already helpful, it is the addition of feedback that makes LLM-based chatbots effective in supporting participants’ goal setting. In contrast, adaptive suggestions were less effective. Altogether, our study shows how to design chatbots by operationalizing psychological frameworks to provide effective support for reaching behavioral goals.
\end{abstract}


\begin{teaserfigure}
    \centering
  \includegraphics[width=0.74\linewidth]{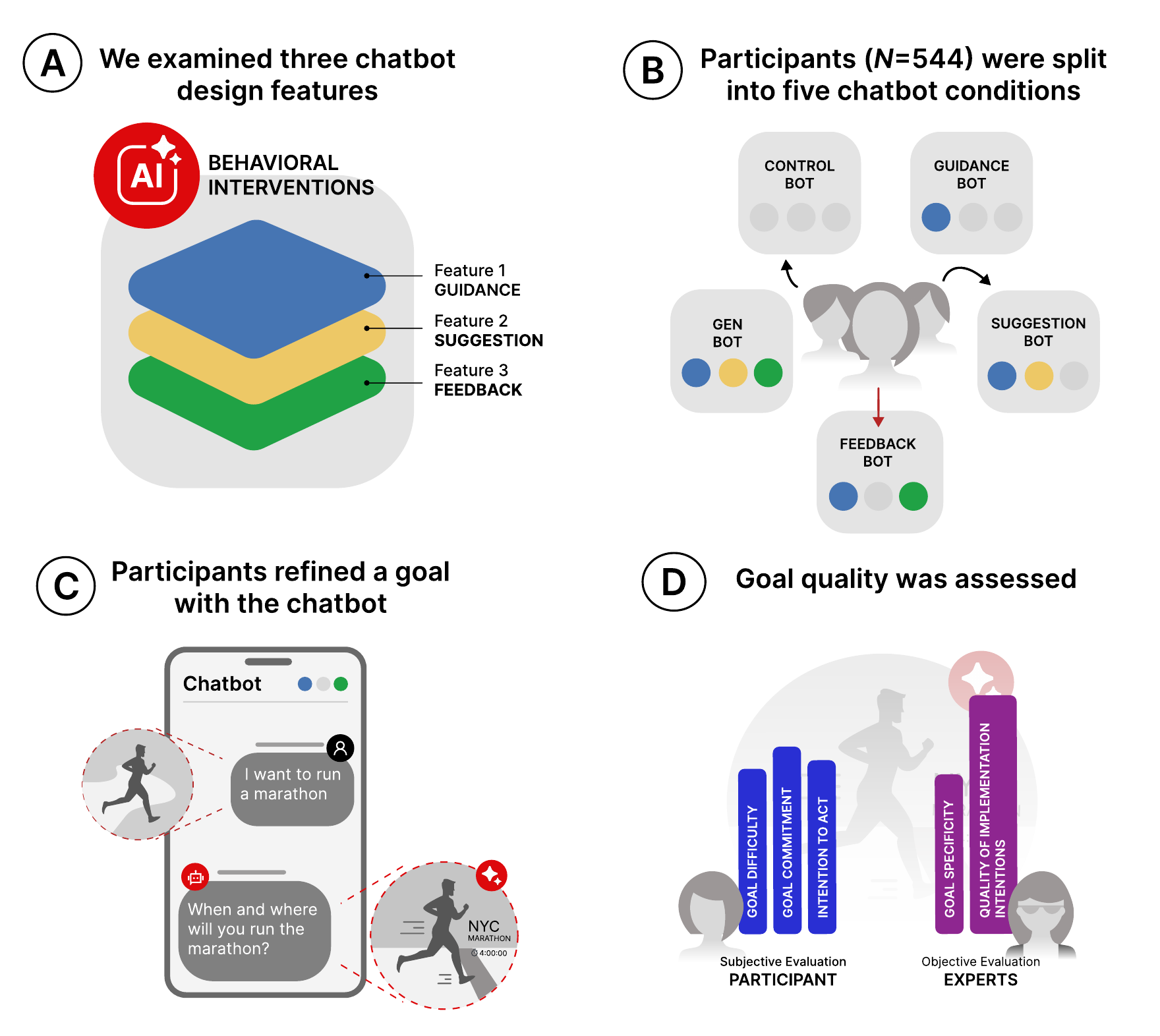}
  \caption{\textbf{Illustration of how different LLM design features can support the application of psychological frameworks}. We \textbf{(A)} examined three features, namely guidance, suggestion, and feedback. To test their effectiveness, these features were \textbf{(B)} incorporated across five different conditions. In \textbf{(C)}, the example of the Feedback Bot is depicted: Participants state their goal and are asked to specify certain details when they are missing. The goals and actions were evaluated across subjective and objective measures \textbf{(D)}.}
  \Description{This figure illustrates the study’s design and evaluation pipeline. It visualizes three behavioral design interventions—guidance, suggestion, and feedback—as layered elements introduced into different chatbot conditions tested with 543 participants. It also shows an example interaction where the chatbot refines a participant’s broad goal (‘I want to run a marathon’) by prompting for specifics (‘When and where will you run it?’), and displays how goal quality was assessed using subjective participant ratings (e.g., difficulty, commitment) and objective expert measures (e.g., specificity, implementation intentions). Together, the panels reveal how intervention features were implemented, how users refined goals in conversation, and how outcomes were evaluated.}
  \label{fig:teaser}
\end{teaserfigure}


\maketitle

\section{Introduction}
\label{sec:introduction}

\new{Many individuals who formulate aspirations, such as becoming healthier, advancing their careers, or finding a partner, struggle to act on them{\textsuperscript{(R2)}}}. Psychological research has identified several levers that help individuals bridge this gap by setting effective goals and action plans. According to goal setting theory, an effective goal should be as specific and challenging as possible while remaining attainable~\citep{locke_theory_1990}. According to implementation intention theory, action plans should specify both a specific cue and a specific action associated with that cue~\cite{gollwitzer_implementation_1999}.
However, even though these theories have been found by academic research to be effective, people often struggle to independently apply these theories in their day-to-day lives when setting personal goals~\citep{armitage_effectiveness_2009, ziegelmann_adoption_2006}. In principle, trained professionals in fields such as coaching, healthcare, and therapy can help people set goals more effectively, yet access to these human resources remains limited~\citep{terblanche_comparing_2022}.

The above challenges present an opportunity for technology, particularly intelligent chatbots, to complement human expertise by serving as accessible digital assistants \cite{zhu_review_goals_chi}.
\new{However, prior to the advent of large language models, such interventions relied on rigid, rule-based architectures that selected responses from predefined options, limiting context-aware interaction and sustained message contingency ~\cite{terblanche_smooth_2024}. 
These constraints restrict an agent's ability to demonstrate \textit{coherence} and \textit{attentiveness}, key constructs for evaluating user experiences with artificial social agents~\cite{Fitrianie2025}. Consequently, such systems might fail to establish a strong \textit{social presence}, limiting their capacity to provide the nuanced, human-like support characteristic of a trained professional~\cite{Hua2025}.
Modern chatbots, empowered by generative AI, could overcome these limitations and empower individuals to more successfully apply established psychological theories in their goal setting processes. Recently, there has been a growing body of HCI research showing the potential of large language models in behavioral \citep{Stade2024} and attitudinal \citep{holbling_meta-analysis_2025} interventions .\textsuperscript{(R1)} }
A key strength of LLMs is their ability to understand and generate human-like text. This capability creates opportunities to adapt psychological interventions from human-delivered to LLM-delivered formats, making them scalable and easily accessible \citep{demszky_using_2023}. As a result, LLM-based chatbots are explored to support various forms of coaching \new{\citep{Arakawa2024,terblanche_exploring_2024}\textsuperscript{(R2)}}. However, to our knowledge, previous studies have not systematically examined which mechanisms in LLM-based interventions most effectively support users in setting specific goals and actionable plans. 

To address the above gap, we aim to study the following research question:

\begin{quote}
\textbf{Research Question:} \textit{To what extent can LLM-based chatbots support users' adoption of goal setting theory and implementation intentions?}
\end{quote}

To answer this question, we conducted a preregistered randomized controlled experiment ($N = 543$) using a between-subjects design. Therein, we tested three different chatbot mechanisms (namely, \textit{guidance}, \textit{suggestions}, and \textit{feedback}), which we incorporate in five chatbot conditions:
\new{(1) a control condition using a basic chatbot to deliver two fixed questions sequentially, functionally mirroring a linear questionnaire without adaptive capabilities\textsuperscript{(2AC)}};
(2) a scripted rule-based chatbot providing step-by-step \textit{guidance} without any intelligent LLM functionality;
(3) a chatbot that builds upon the rule-based system by additionally providing LLM-generated \textit{suggestions} alongside the core \textit{guidance};
(4) a personalized feedback chatbot that offers \textit{guidance} supplemented with personalized LLM-generated \textit{feedback}; and
(5) a fully interactive LLM-based chatbot that combines all three mechanisms: \textit{guidance}, \textit{suggestions}, and \textit{feedback}.
Participants were asked to set personally meaningful goals with one of the five chatbots through a step-by-step process designed to help them better formulate these goals and develop appropriate action plans. Participants' goals ranged from ``running a marathon'' and ``getting a promotion'' to ``learning Python'' and ``getting a good grade in a university course''.

We evaluated participants’ adoption of goal setting theory and implementation intentions by examining the components that constitute effective goals and actions according to both theories, measuring them independently for each participant. Informed by goal setting theory, we rated the specificity of goals. Participants also rated goal difficulty and their commitment towards the goal. Finally, we rated the quality of the implementation intentions and also asked the participants to rate their intention to act on the implementation intentions they had formed.
We find that guidance and feedback in LLM-based chatbots help participants set higher quality goals and implementation intentions but do not increase goal commitment or the intention to act on them. Meanwhile, we observe no effects of the suggestion feature on primary goal outcomes.

Our work contributes to both theoretical understanding and practical design of AI-based behavioral interventions. Theoretically, we illustrate how individuals can be effectively supported in setting evidence-based goals and action plans. Moreover, by explicitly focusing on the underlying mechanisms through which LLMs influence user behavior, our work advances HCI research, moving beyond outcome-focused evaluations to explore foundational processes in human–chatbot interaction.
From a practical standpoint, our work demonstrates how LLMs can offer a scalable, easily accessible, and cost-effective solution for helping individuals set more effective goals and action plans, grounded in well-established psychological theory. This opens opportunities across diverse domains—including organizational management~\cite{locke_building_2002}, sports~\cite{Williamson2022}, healthcare~\cite{Bodenheimer2009}, and personal development~\cite{MacLeod2007}—where goal setting is a critical component of success. 

\section{Related Work}
\label{sec:relatedwork}
\subsection{Theoretical Foundations of Goal Attainment}
The goal attainment process can be distinguished into two fundamental stages~\citep{thurmer_planning_2017}. The initial stage involves setting goals effectively and committing to them, while the subsequent stage focuses on translating these goals into specific, actionable steps, a process often described as goal striving~\citep{lewin_vorsatz_1926, lewin_level_1944, heckhausen_thought_1987, gollwitzer_action_1990}. 

\textit{Goal setting theory }primarily addresses the first stage, emphasizing that specific and challenging yet attainable goals significantly enhance performance and facilitate greater goal achievement~\citep{locke_theory_1990} \new{(see Table~\ref{tab:research-streams})}. Additionally, this theory differentiates between learning and performance goals. Learning goals, which aim to develop strategies and acquire knowledge, enhance performance on novel or complex tasks. Conversely, performance goals, oriented towards achieving specific outcomes, become particularly effective once individuals have acquired the necessary competencies~\citep{locke_building_2002}.

\begin{table*}[htbp]
\centering
\caption{\new{Research Gaps in Goal Setting Research.}}
\label{tab:research-streams}
{\footnotesize\color{black}

\begin{tabularx}{\textwidth}{p{0.16\textwidth}p{0.50\textwidth}X}
\toprule
\textbf{Research stream} & \textbf{What prior work shows} & \textbf{Open questions} \\
\midrule
\raggedright\textbf{Empirical evaluation of goal setting theory and implementation intentions} &
Meta-analytical evidence shows that goal setting and implementation intentions are effective across domains and behaviors \cite{epton_unique_2017, gollwitzer_implementation_2006, sheeran_when_2024}, but individuals often struggle to formulate specific and challenging goals without external support \cite{armitage_effectiveness_2009, clarke_goal-setting_2009, smith_contribution_2013, ziegelmann_adoption_2006}. &
How can individuals be supported in the process of setting their individuals goals? \\
\midrule
\raggedright\textbf{Pre-LLM technology for goal setting} &
HCI research has extensively examined technologies for goal setting \cite{zhu_review_goals_chi}. While some work focuses on the goal-setting process itself, most studies combine goal specification with support for goal attainment (e.g., monitoring, feedback, visualization). Prior work spans a range of system types, including:
\begin{itemize}
\item \textbf{Social and peer support systems}: 
Crowdsourced generation of personalized action plans\cite{agapie_plansourcing_2016, agapie_crowdsourcing_2018}.
\item \textbf{Mobile and app-based interventions}: Support goal setting and goal attainment through self-defined goals, self-monitoring, and feedback mechanisms such as rewards or social sharing. \cite{munson_exploring_2012, consolvo_goal-setting_2009, murnane_narrative-based_2023}.
\item \textbf{Wearables}: Continuous sensing and real-time feedback on goal progress \cite{kim_towards_2020, fritz_persuasive_2014}.
\item \textbf{Ambient and tangible interfaces}: Physical, environmental, or shared-display visualizations of goal state or progress\cite{sauve_econundrum_2020, botros_go_2016, denefleh_sensorstation_2019}.

\end{itemize} &
How can LLMs enhance existing technologies in supporting individuals with setting goals in a domain-agnostic way?  \\
\midrule
\raggedright\textbf{LLM based support for goal setting} &
LLMs have been explored in HCI through proof-of-concept systems and experimental evaluations across diverse goal-setting domains. However, the LLM-specific mechanisms underlying their effectiveness remain underexplored.
\begin{itemize}
\item \textbf{Proof-of-concept systems and artifacts}:
LLM-based systems integrating sensor and wearable data to support goal setting and goal attainment through natural language interaction \cite{jorke_gptcoach_2025, wu_mindshift_2024, nepal_mindscape_2024, xu_goals_2025}. Recent work further extends this approach through multi-agent LLM architectures with specialized roles \cite{heydari_anatomy_2025, wang_exploring_2025}.
\item \textbf{Experimental evaluation of LLMs for goal setting}: Experimental evaluation of LLM-based support for the goal-setting process, reporting goal attainment rates comparable to human coaches\cite{terblanche_comparing_2022, terblanche_adoption_2022}. 
\end{itemize}
 & 
What are the LLM-specific design features that support individuals in setting goals beyond earlier rule-based chatbots? \\
\bottomrule
\end{tabularx}

}
\Description{A three-row table mapping related work to open research questions. It contrasts traditional goal-setting theory and pre-LLM technologies (such as apps, wearables, and crowd-sourcing) against emerging proof-of-concept LLM systems. The table highlights that while goal setting is effective, users struggle to do it alone, and raises questions about how LLMs and multi-agent systems can specifically enhance this process beyond previous rule-based technologies.}
\end{table*}

In the second stage, which refers to action execution, \citet{gollwitzer_implementation_1999} recommends setting \textit{implementation intentions}. Implementation intentions are structured, actionable plans formulated in an if-then format, explicitly linking specific situational cues to predefined actions. These intentions clearly specify when, where, and how an action should be executed (\emph{``If I encounter situation A, I will perform action B''}), thereby increasing the likelihood that the intended action is activated automatically when the goal-relevant situation arises. This is in contrast to goal intentions, which merely state a general commitment to achieving a certain outcome (``I want to achieve B'') without specifying the concrete conditions for action execution.

There is strong evidence suggesting that individuals may require support in formulating their goals and actions~\cite{clarke_goal-setting_2009, smith_contribution_2013,armitage_effectiveness_2009, ziegelmann_adoption_2006}. This need for support has been a long-standing focus in HCI, where researchers have explored technologies for personal informatics, behavior change, and goal setting support~\cite{zhu_review_goals_chi}. For example, in clinical contexts, several procedures have been developed to assist clinicians and patients in defining rehabilitation goals~\citep{clarke_collaborative_2006, cohen_psychiatric_1991, kiresuk_goal_1968}. Studies in this field have demonstrated that training and support in goal setting can improve the goal quality~\citep{clarke_goal-setting_2009, smith_contribution_2013}. 
Moreover, evidence suggests that individuals may form ineffective implementation intentions. Accordingly,
self-generated implementation intentions are less effective than those formulated with professional guidance~\citep{armitage_effectiveness_2009, ziegelmann_adoption_2006}. Consequently, structured guidance in developing evidence-based goals and action plans may be necessary. \new{In their meta-analysis, \citet{epton_unique_2017} highlighted the critical gap in understanding how to best support individuals during goal formulation (see first row Table~\ref{tab:research-streams}), stating that ``it is not clear how people should be encouraged to set goals; for example, whether simply assigning goals is sufficient or if targeted training in goal formulation is necessary, and what role emerging technologies might play in supporting effective goal setting.''}

\subsection{\new{Technology for Behavioral Interventions}}
HCI has a long tradition of designing digital interventions to promote behavior change \cite{ekhtiar_goals_2023, zhu_review_goals_chi}. Prior work has explored how technologies across diverse streams (see Table~\ref{tab:research-streams})—ranging from social support systems~\cite{agapie_plansourcing_2016} and ambient interfaces~\cite{sauve_econundrum_2020} to mobile apps \cite{MilneIves2023} and wearables \cite{DelValleSoto2024}—can support people in pursuing their goals. For example, \citet{Hiniker2016} developed the \textit{MyTime} app, combining goal setting, self-monitoring, and time-limit nudges to reduce the time spent on self-identified waste-of-time applications. While these systems can deliver elements of guidance and feedback (e.g., via pop-up notifications or visualized progress), they typically rely on pre-defined rules, static templates, or "Wizard of Oz" approaches~\cite{agapie_plansourcing_2016}. 
This creates a trade-off: rule-based systems are scalable but often rigid and unable to process the semantic nuance of open-ended user goals, whereas human-in-the-loop systems offer personalization but are difficult to scale. Furthermore, \citet{zhu_review_goals_chi} highlight in their systematic review that behavioral interventions overwhelmingly target single domains, particularly health. 
However, human goals extend well beyond these specific domains. Accordingly, there remains a lack of systematic support within HCI for agnostic tools supporting also multi-domain, social, or qualitative goals driven by high intrinsic motivation \cite{zhu_review_goals_chi}.

At the same time, people are increasingly turning to general-purpose LLMs as informal therapists or coaches, discussing problems and seeking advice \cite{jung__2025}. Yet, \new{out of the box domain agnostic systems like ChatGPT~\cite{chatgpt}\textsuperscript{(R2)}} do not embed established scientific principles or frameworks in systematic manner \cite{Lawrence2024}. The same tendency is evident in HCI research. A recent meta-analysis reported that while most papers mention theories, only a small fraction of them use them to inform technology design~\cite{zhu_review_goals_chi}. This raises concerns about reliability in high-stakes contexts and decreases the effectiveness of these interventions.

Chatbots have also been specifically applied in behavior change interventions, although their specific contributions to goal quality remain understudied. For example, in the context of mental health, \citet{sharma2024} deployed an LLM-based tool, supporting self-guided cognitive restructuring, while others have aimed to reduce problematic smartphone use~\cite{Olson2022}. In goal setting, \citet{terblanche_comparing_2022} showed that chatbot-supported participants achieved goal-attainment levels comparable to those supported by human coaches, both outperforming participants who received only informational materials. However, LLM-based chatbots did not outperform rule-based chatbots in facilitating goal setting related to technology adoption \cite{terblanche_smooth_2024}. Existing research primarily evaluates overall intervention effectiveness, often neglecting underlying mechanisms \cite{liu_investigating_2024}. This ``black-box'' approach restricts theoretical understanding, making it difficult to understand the heterogeneous results \cite{balan_use_2024}. This presents a design challenge for HCI: \new{To advance HCI and, specifically, the research field of behavior change interventions, there is a need for empirical evaluation of design mechanisms that support behavior change interventions.}

\new{In summary, despite evidence that individuals struggle to formulate effective goals and implementation intentions, no prior work has systematically examined to what extent specific design mechanisms in LLM-based chatbots can support effective goal setting beyond prior technological support such as rule-based chatbots.} To address this gap, we investigate three key characteristics—guidance, suggestions, and feedback—that may play a central role in supporting effective goal setting and action planning.

\section{Hypotheses: Design Features in Behavioral Interventions}\label{sec:hypothesis}
Chatbots can provide \textit{guidance}\label{back:guidance} throughout interventions, thereby offering structured, incremental support rather than delivering all instructions at once. Evidence demonstrates guided chatbot interventions are effective for psychotherapy in depression, anxiety~\citep{lim_chatbot-delivered_2022}, stress, and acrophobia~\citep{abd-alrazaq_effectiveness_2020}—even with less sophisticated rule-based or retrieval-based chatbots predating large language models. For goal setting specifically, guidance alone appears valuable. For example, \citet{terblanche_comparing_2022} showed that a chatbot delivering a step-by-step goal setting intervention outperformed providing participants with information alone. Thus, we predict:

\begin{description}
    \item [H1:] \textit{\textbf{Guidance} increases the a) goal difficulty, b) specificity and c) quality of implementation intentions.}
\end{description}

Beyond pure guidance, LLM-based chatbots can provide adaptive \textit{suggestions}\label{back:suggestions}, generating \new{ personalized examples \textsuperscript{(R1)}} tailored to the ongoing conversation. Adaptivity is a key differentiator: unlike static examples, LLMs can incorporate prior user input and context to tailor suggestions to individual circumstances. Prior work shows that adaptivity in chatbot interactions supports effective behavioral interventions \cite{heintzelman_personalizing_2023, jeong_chatbot_2023, liu_investigating_2024, shumanov_making_2021}. Additionally, LLM-generated suggestions enhance output quality in creation tasks \cite{anderson_homogenization_2024, doshi_generative_2024} or self-guided mental-health interventions \cite{sharma2024}. Applied to goal setting, providing users with theory-informed examples tailored to their personal situation may encourage them to articulate goals and actions that are more specific and ambitious than those formulated independently.  

Therefore, we hypothesize that adaptive suggestions enhance goal quality in the goal setting context:

\begin{description}
    \item [H2:] \textit{\textbf{Suggestions} increase the a) goal difficulty, b) goal specificity and c) quality of implementation intentions.}
\end{description}

Additionally, LLM-based chatbots can utilize \textit{feedback}\label{back:feedback} within interventions—a collaborative process providing real-time \new{ interactive{\textsuperscript{(R1)}}} recommendations to enhance their goals or actions. \new{Unlike suggestions, which proactively offer examples before users formulate their responses, feedback reactively evaluates and critiques user-generated content after initial formulation.{\textsuperscript{(R2)}}} \citet{liu_investigating_2024} demonstrated, for example, that positive psychology interventions achieved greater effectiveness when chatbots employed an interactive question-and-answer approach, adapting to user input and offering real-time feedback. In the goal setting context, feedback enables LLM-based chatbots to detect when essential components of theory-based formulations are missing or suboptimal and to prompt immediate improvements. For example, if a participant specifies an implementation intention with a location unlikely to trigger action, the chatbot can highlight this weakness and ask for improvement. We thus hypothesize: 

\begin{description}
    \item [H3:] \textit{\textbf{Feedback} increases the a) goal difficulty, b) goal specificity, and c) quality of implementation intentions.}
\end{description}

In addition, we predict that the combination of all three features may create synergistic effects and surpass the impact of each feature alone. For example, when the chatbot provides feedback, adaptive suggestions can support users in implementing this feedback more effectively. Thus, we hypothesize:

\begin{description}
    \item [H4:] \textit{The combination of all three design features---\textbf{guidance, suggestions, and feedback}--- surpasses each feature alone regarding a) goal difficulty, b) goal specificity, and c) quality of implementation intentions.}
\end{description}

Beyond technological features, psychological characteristics inherent in LLM-based chatbots may shape intervention effectiveness. A particularly relevant mechanism is social presence, defined as the feeling of interacting with another human-like entity \cite{biocca2003toward}. LLM-based chatbots, with their ability to guide users, adapt suggestions, and offer feedback, can create stronger perceived social presence \citep{go_humanizing_2019, janson2023leverage}. 
\new{Theoretically, these capabilities foster social presence through \textit{message contingency}. Defined as the extent to which a message relates to and builds upon preceding utterances, contingency is a primary driver of social presence in human interaction \citep{rafaeli1988interactivity}. Unlike more rigid, rule-based systems, LLMs generate responses that are highly contingent on the user's specific input. Drawing on the Computers Are Social Actors (CASA) paradigm \citep{nass1994computers, reeves1996media}, such contingent interactivity serves as a social cue. When a system demonstrates that it is "listening" by dynamically adapting its output, it triggers a heuristic response where users attribute a psychological state to the system, thereby heightening perceived social presence \citep{sundar2016theoretical}.\textsuperscript{(R1)}}
From a motivational perspective, social presence can strengthen engagement and accountability by simulating interpersonal goal setting situations. Previous research in social psychology highlights that goal commitment increases when goals are set socially rather than privately, meaning that making goals public increases accountability and thus strengthens commitment \citep{epton_unique_2017, ingledew_work-related_2005, klein_goal_1999}. \citet{terblanche_comparing_2022} further demonstrated that participants interacting with AI-chatbots achieved higher goal attainment compared to those receiving only static information, thereby underscoring the value of human-like conversational qualities.
In our intervention, we therefore expect that participants engaging with an LLM-based chatbot will experience increased social presence, which in turn will foster stronger goal commitment. This reflects an indirect effect, such that the type of chatbot influences perceived social presence, which in turn predicts participants’ goal commitment.

\begin{description}
    \item [H5:] \textit{\textbf{Social presence mediates} the relationship between chatbot type (GenBot vs. ControlBot) and \textbf{goal commitment}.}
\end{description}

To our knowledge, no existing research explicitly investigates the specific chatbot characteristics—guidance, suggestions, feedback, and their interplay in the context of psychological interventions. Furthermore how these features influence the social presence of a chatbot and if that might enhance the quality and effectiveness of goal setting interventions, is yet to be answered

\section{Methodology}
\label{sec:methodology}

\new{To systematically analyze the effectiveness of chatbot features in goal setting and implementation intention formation, we conducted a preregistered\footnote{\url{https://osf.io/kjmdw/overview?view_only=ec76774bd89b4c5e929ade14375f27d4}}\textsuperscript{(R1)}} randomized controlled online experiment between 17 and 22 December 2024. In the study, participants formulated a personally relevant goal and implementation intention in collaboration with one of five chatbots. 
In the following, we describe our prototype design (Section~\ref{met:design}), intervention groups (Section~\ref{met:groups}), experimental procedure (Section~\ref{met:procedure}), study measures (Section~\ref{subsec:measures}), study population (Section~\ref{met:population}), statistical analysis (Section~\ref{met:statistics}), and ethical considerations (Section~\ref{met:ethics}).

\subsection{Prototype Design}\label{met:design}
\begin{figure*}[t]
    \centering
    \includegraphics[width=1\linewidth]{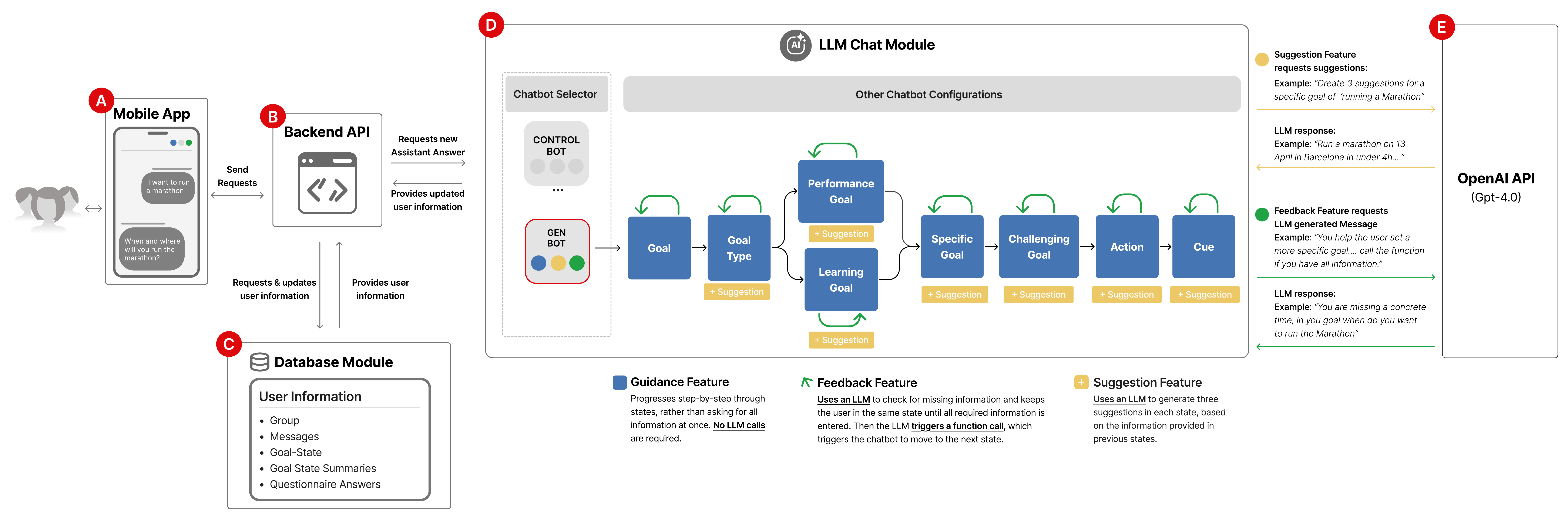}
    \caption{\textbf{System architecture of the goal setting chatbot application.} \textbf{(A)} The mobile app serves as the user interface, enabling participants to interact with the chatbot. \textbf{(B)} Requests are routed through the backend API, which handles communication between the app, database, and chatbot module. \textbf{(C)} The database module stores user information (e.g., messages, goal states) and manages group assignment and questionnaire tracking. \textbf{(D)} The LLM chat module implements a state-based architecture for structured goal setting conversations. Each conversational state corresponds to a step in the procedure. \textbf{(E)} The system connects to the OpenAI API for LLM-generated outputs.}
    \Description{This architecture diagram depicts how the mobile app interface, backend API, and database module interact to support structured goal conversations with an LLM-based chat module. User input from the mobile app is routed via the API, which accesses stored user data—such as goal state and messages—and feeds structured conversation states (goal, goal type, specific goal, challenging goal, action, cue) to the LLM chat module. Interventions (guidance, suggestion, feedback) are integrated into the state transitions. The final panel shows the connection to the OpenAI API, which generates the chatbot’s responses.}
    \label{fig:states}
\Description{This architecture diagram depicts how the mobile app interface, backend API, and database module interact to support structured goal conversations with an LLM-based chat module. User input from the mobile app is routed via the API, which accesses stored user data—such as goal state and messages—and feeds structured conversation states (goal, goal type, specific goal, challenging goal, action, cue) to the LLM chat module. Interventions (guidance, suggestion, feedback) are integrated into the state transitions. The final panel shows connection to the OpenAI API, which generates the chatbot’s responses.}
\end{figure*}

Since mobile technologies have been shown to deliver behavioral interventions more effectively than traditional or desktop-based methods~\cite{Lindhiem2015}, we used a dedicated iOS-based platform, previously developed by the research group for this study. The platform was implemented using Flutter, with a Python-based backend server to handle data storage and LLM interactions.  User authentication and data storage were handled via Google Firebase.

 To maintain consistency across extended conversations of over 30 turns, we designed a state-based chatbot architecture (see Figure~\ref{fig:states}). This approach addresses two key challenges with current LLMs: their tendency to drift during lengthy structured conversations, and the difficulty of managing long, complex and multi-step processes through system prompts alone. Our state-based design maps each step of the goal setting procedure to a specific conversational state, ensuring the chatbot maintains focus throughout the interaction. Depending on the experimental condition (see Section~\ref{met:groups}), responses were generated either through static scripted messages or dynamically via the LLM, with function calling enabling state transitions. \new{To support this architecture, each conversational state had an associated function that the LLM could invoke when it determined that the user had completed the step. The backend verified the call and then advanced the conversation to the next state. In practice, with all participants reaching the end state, this mechanism was reliable across extended dialogues.\textsuperscript{(2AC)}}
 
 \new{The prototype underwent more than ten iterative development cycles, each tested with university students and small batches of Prolific participants who provided qualitative feedback. This feedback informed prompt refinement, the development of the three chatbot features, and the verification that the experimental conditions where stable and differed solely in this targeted aspect.{\textsuperscript{(2AC)}}}
 The implementation of the LLM backend is available in our project repository\footnote{https://github.com/michel-schimpf/goal-setting-llm-backend.git}.

The LLM used in this study was \textit{gpt-4o-2024-11-20}, with temperature set to $0.1$. \new{We selected this specific model because, at the time of the study, it was the most performant option available and the only model capable of reliably managing the required state transitions. We note, however, that newer open-source models released in 2025, such as Mistral Medium 3.1~\cite{mistral2025medium} and DeepSeek-R1~\cite{deepseek2025r1} now possess the necessary function calling reliability to control the state architecture.\textsuperscript{(R1)}} We utilized a low temperature setting, because we found that higher temperatures sometimes caused the LLMs to diverge from the desired behavior. Furthermore, the lower temperature made it easier to test the application and replicate results.

\begin{figure*}[t]
    \centering
    \includegraphics[width=1\linewidth]{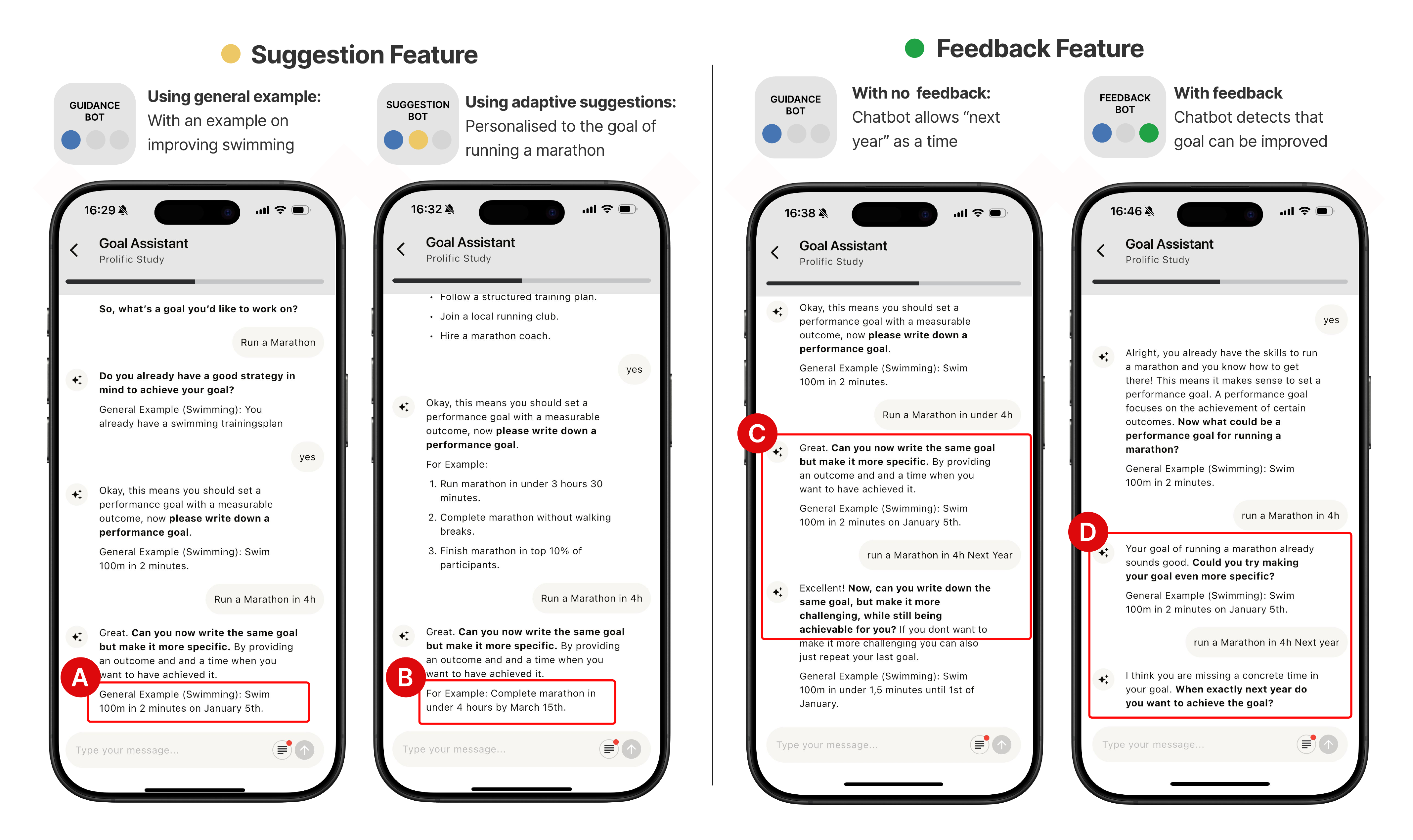}
    \caption{\textbf{Example interactions with the goal assistant chatbots demonstrating different design features.} \textbf{(A)} Guidance with a general example: The chatbot uses a general swimming example to help the user think about learning-oriented goals. \textbf{(B)} Suggestions with personalization: Adaptive strategies tailored to a marathon goal are provided. \textbf{(C)} No feedback: The chatbot accepts an underspecified cue (“this evening”) without prompting for clarification. \textbf{(D)} Feedback: The \textit{FeedbackBot} detects that the cue is vague and encourages the user to make it more concrete and situational.}
    \Description{This figure compares four conversational snapshots from the Goal Assistant chatbots. In the first, the GuidanceBot uses a general example (swimming) to guide the user toward setting a learning-oriented goal. In the second, the SuggestionBot adapts to a marathon goal, offering personalized strategy suggestions. The third snapshot shows a bot with no feedback, which accepts a vague goal (‘run a marathon in four hours next year’) without prompting for refinement. The fourth shows the FeedbackBot identifying the vague time frame and prompting the user to make the goal more concrete by specifying when the goal will be achieved.}
    \label{fig:chatbots}
\end{figure*}

\subsection{Intervention Groups}\label{met:groups}
We developed five distinct intervention groups using 5 different chatbots representing progressive levels of conversational intelligence and support. Each chatbot helped users set a personally meaningful goal and incorporates different combinations of three core features: \textit{guidance}, \textit{suggestions}, and \textit{feedback}.

The \textit{ControlBot} (\controlbotbadge) serves as the baseline condition, implementing none of the three conversational features. Instead, it functions as a static input form. Participants receive two standardized messages: the first outlines the principles of effective goals based on goal setting theory, and the second describes the characteristics of effective implementation intentions. Participants then enter their responses without any interactive assistance. This condition mirrors the \new{capabilities of filling out a static form, while maintaining the chatbot interaction modality used in the other conditions and therefore provides a benchmark for evaluating the added value of of the conversational guidance feature.{\textsuperscript{(R1)}}}

The \textit{GuidanceBot} (\guidancbotbadge) extends the control condition by providing step-by-step \textit{guidance}, directing participants through a predefined sequence for formulating goals and corresponding implementation intentions. It employs rule-based interactions to guide users systematically through key steps such as selecting a goal type (learning vs. performance), specifying the goal, calibrating its difficulty, identifying actions, and establishing cues. Unlike LLM-based systems, the GuidanceBot does not rely on generative models; instead, it delivers static messages and advances to the next state once a user responds. As such, it represents a form of chatbot that could have been implemented prior to the advent of generative AI. This design allows us to isolate and evaluate the added benefits of LLM-driven guidance beyond static, rule-based interaction.

The \textit{SuggestionBot} (\suggestionbotbadge) builds on the structured \textit{guidance} approach by adding adaptive \textit{suggestions} generated by an LLM. During the interaction, the system summarizes user input and provides three contextually relevant examples at each step to help participants formulate their responses. For instance, if a user indicates an interest in marathon running, the chatbot generates examples specifically related to marathon training goals (see Figure~\ref{fig:chatbots}, panels A and B). These adaptive suggestions are appended to the same static \textit{GuidanceBot} message and replace the static examples used there, while all other components of the interaction remain identical. This design isolates the effect of adaptive suggestions, allowing a direct comparison between static and LLM-generated support.

\begin{figure*}[t]
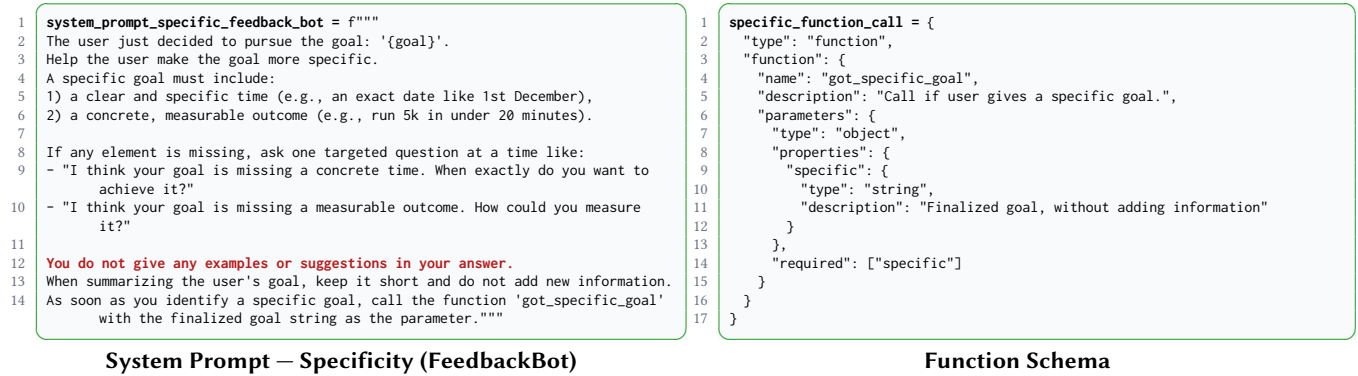

\centering

\begin{minipage}[t]{0.50\textwidth}
\begin{codecardGreen}[title=System Prompt — Specificity (FeedbackBot)]
§system_prompt_specific_feedback_bot = § f"""
The user just decided to pursue the goal: '{goal}'.
Help the user make the goal more specific.
A specific goal must include:
1) a clear and specific time (e.g., an exact date like 1st December),
2) a concrete, measurable outcome (e.g., run 5k in under 20 minutes).

If any element is missing, ask one targeted question at a time like:
- "I think your goal is missing a concrete time. When exactly do you want to achieve it?"
- "I think your goal is missing a measurable outcome. How could you measure it?"

@You do not give any examples or suggestions in your answer.@
When summarizing the user's goal, keep it short and do not add new information.
As soon as you identify a specific goal, call the function 'got_specific_goal' with the finalized goal string as the parameter."""
\end{codecardGreen}
\end{minipage}\hfill
\begin{minipage}[t]{0.49\textwidth}
\begin{codecardGreen}[title=Function Schema,]
§specific_function_call = §{
  "type": "function",
  "function": {
    "name": "got_specific_goal",
    "description": "Call if user gives a specific goal.",
    "parameters": {
      "type": "object",
      "properties": {
        "specific": {
          "type": "string",
          "description": "Finalized goal, without adding information"
        }
      },
      "required": ["specific"]
    }
  }
}
\end{codecardGreen}
\end{minipage}

\caption{\textbf{Example of a system prompt (left) and function call (right) for the ''specific goal'' state of the \textit{FeedbackBot}.} Each state in the \textit{FeedbackBot} has a similar prompt, with an explicit instruction not to give suggestions, ensuring it is differentiated from the SuggestionBot condition. This instruction is removed in the GenBot.}
\Description{This figure displays the prompt structure and associated function logic used in the ‘specific goal’ conversational state of the FeedbackBot. The system prompt instructs the chatbot to help users refine their goal by ensuring it includes a clear time (e.g., ‘1st December’) and measurable outcome (e.g., ‘run 5k in under 20 minutes’), and to ask targeted clarifying questions if either element is missing—without offering examples or suggestions. The adjacent function schema shows that once a specific goal is identified, the chatbot executes a function call that captures and passes the finalized goal string for further processing.}
\label{fig:prompt_function}
\end{figure*}

The \textit{FeedbackBot} (\feedbackbotbadge) combines the structured \textit{guidance} framework with an iterative \textit{feedback} mechanism powered by an LLM. Unlike the \textit{GuidanceBot}, the system does not immediately advance after each user response. Instead, it keeps participants in a given state until all required information is provided, with the LLM generating clarifying follow-up questions when essential elements are missing (e.g., “Your goal seems to lack a concrete time frame. When exactly do you want to achieve it?”) (see Figure~\ref{fig:chatbots}, panels C and D).  All messages in this condition are generated by the LLM using a system prompt similar to the one shown in Figure~\ref{fig:prompt_function}. Once sufficient information has been gathered, the LLM issues a function call that triggers the transition to the next state.

Finally, the \textit{GenBot} (\genbotbadge{}) represents the most comprehensive condition, combining \textit{guidance}, \textit{suggestions}, and \textit{feedback} into a fully featured conversational system. This condition reflects the capabilities of a modern LLM-based goal setting assistant without artificial restrictions, offering structured support, context-specific examples, and iterative refinement throughout the goal formulation process. Implementation follows the approach of the \textit{FeedbackBot}, but additionally provides adaptive suggestions at the start of each state (as in the \textit{SuggestionBot}) and can generate further suggestions within the feedback loop when appropriate.

\subsection{Procedure}\label{met:procedure}

Participants were recruited via Prolific and required to use a mobile device to access the study application. After downloading the application, they received an information sheet and provided informed consent.
Upon consenting, each participant was randomly assigned to one of the five experimental conditions (see Section~\ref{met:groups}).

Participants interacted with the chatbot according to their assigned condition. Across all conditions, they were guided to formulate a personally relevant goal and generate a corresponding implementation intention. The interaction was completed entirely on their own device and typically lasted 10–15 minutes.

Immediately after the chatbot session, participants completed a post-study questionnaire, where items were presented on a 5-point Likert scale (1 = “Strongly disagree”; 5 = “Strongly agree”). The questionnaire also included manipulation checks and attention checks. Details of the measured constructs are provided in Section~\ref{subsec:measures}.

\new{To create a consistent sense of accountability across conditions, all participants received the same follow-up question regarding the future evaluation of their goal progress after creating their goal. This question served solely as a general accountability cue and was not used as an experimental manipulation. Participants were debriefed at the end of the study.{\textsuperscript{(2AC)}}}

\subsection{Measures}\label{subsec:measures}

The study relied on two types of measures: (a) self-report questionnaires completed by participants and (b) expert-coded ratings of participants’ written goal and implementation-intention responses. Confirmatory outcomes were selected to test the hypotheses (H1--H4), while exploratory measures provided additional descriptive insights into goal setting, implementation intentions, and social or technological factors.  \new{The complete set of items is presented in Appendix~\ref{appendix:measures}.\textsuperscript{(R1)}}

To test our confirmatory outcomes (H1--H3), which examined whether \textit{guidance}, \textit{suggestions}, and \textit{feedback} improved goal formulation and planning, we measured:  
\begin{itemize}
    \item \textbf{\textit{Goal specificity}}, which was coded on timeframe and measurability (0–2 each, summed to 0–4), with ratings by two independent coders showing excellent agreement (ICC(3,2) = .91).  
    \item \textbf{\textit{Goal difficulty}}, which was assessed using four items from \citet{lee_exploring_1992} ($\alpha = .85$), for example via ''How difficult do you perceive this goal to be?''  
    \item \textbf{\textit{Implementation-intention quality}}, which was coded for the presence of a cue, action, explicit cue–action link, and specification of time/place (0–4 total), again with high inter-rater reliability (ICC(3,2) = .91).  
\end{itemize}

To test H4, which predicted that social presence would mediate effects on goal commitment, we measured:  
\begin{itemize}
    \item \textbf{\textit{Perceived social presence}}, which was captured with four items adapted from~\citet{kumar_research_2006} such as “Regarding the goal assistant, I had a sense of human contact,” rated on a 7-point Likert scale.  
    \item \textbf{\textit{Goal commitment}}, which was measured with five items adapted from \citet{hollenbeck_empirical_1989}, for example, “I am strongly committed to pursuing this goal.”  
\end{itemize}

In addition to these main outcomes, we assessed exploratory measures to provide further insight:  
\begin{itemize}
    \item \textbf{\textit{Behavioral intention}}, which was measured with a single item adapted from \citet{fishbein_predicting_2011}, for example, “I intend to [action].”  
    \item \textbf{\textit{Goal attainability}}, which was measured with a single item (“How attainable do you perceive the goal you just set?”).  
    \item \textbf{\textit{Technology-adoption correlates}}, which were captured using UTAUT2 subscales \citep{venkatesh_consumer_2012}: performance expectancy (5 items), effort expectancy (6 items), hedonic motivation (3 items), and attitudes toward the chatbot (5 items).  
\end{itemize}
Finally, we included self-constructed manipulation checks to confirm whether participants perceived the chatbot as providing guidance, suggestions, or feedback.

\subsection{Study Population}\label{met:population}
 An a priori power analysis revealed a required sample size of \textit{N} = 540 to detect a medium effect size of \textit{f} = 0.15 with a power of $1-\beta = 0.8$ and $\alpha = 0.05$. A total of $N = 576$ participants were recruited in the UK via the platform Prolific. Eligibility criteria for participation included a minimum age of 18, proficiency in the English language, and access to a mobile device to use the application. \new{Here, 33 participants were excluded from the final analysis because they failed at least one attention check (6 failed only the first, 13 only the second, and 14 both), resulting in a final sample of $N = 543$\textsuperscript{(2AC)}}. The final sample consisted of participants aged 18 to 74 years (\textit{M} = 34.35, \textit{SD} = 11.22; 50.27\% female, 49.72\% male). The distribution of demographic characteristics was comparable across experimental conditions (see Appendix~\ref{sec:demographics}).

\subsection{Statistical Analysis}\label{met:statistics}
Analyses were conducted using \textit{R} 4.4.1 \citep{r_core_team_r_2023}. All hypotheses were tested using a significance level of $\alpha = .05$. As assumptions for the preregistered ANOVA analyses were not satisfied,
the Kruskal–Wallis test offered a robust and well-suited alternative for analyzing group differences \cite{KruskalWallis1952}. To assess specific hypotheses, we followed with a Dunn’s post-hoc comparisons \citep{dunn_multiple_1964}, using Holm-adjusted $p$-values \citep{holm_simple_1979} to control for multiple testing. We conducted the mediation analysis using the PROCESS macro in R \citep{hayes_introduction_2022}. For assessing the indirect effect, a bootstrap mediation analysis (20,000 resamples) was used.

\subsection{Ethics}\label{met:ethics}
The study was conducted in accordance with ethical guidelines and approved by the Ethics Committee of the Department of Engineering at the University of Cambridge. All participants provided informed consent prior to participation. Participants were informed about the voluntary nature of their participation, their right to withdraw at any time without penalty, and the anonymized handling of their data. No personally identifiable information was collected, and all responses were stored in accordance with GDPR guidelines. Participants were compensated via the Prolific platform in accordance with its fair pay policy.

\FloatBarrier

\begin{table*}[t]
\caption{\textbf{Descriptive statistics for all study outcomes across the five chatbot conditions (ControlBot, GuidanceBot, SuggestionBot, FeedbackBot, and GenBot).} Reported are means ($M$) and standard deviations ($SD$) for goal-related outcomes, perception outcomes, and technology adoption outcomes. Sample sizes are shown below each condition label.}
\label{tab:descriptives_by_condition}
\centering
\small 

\begin{tabular}{@{}l*{10}{c}@{}}
    \toprule
    & \multicolumn{2}{c}{\makecell{\controlbotbadge \\ \textbf{ControlBot} \\[-2pt] \scriptsize(\textit{n}=111)}}
    & \multicolumn{2}{c}{\makecell{\guidancbotbadge \\ \textbf{GuidanceBot} \\[-2pt] \scriptsize(\textit{n}=108)}}
    & \multicolumn{2}{c}{\makecell{\suggestionbotbadge\\ \textbf{SuggestionBot} \\[-2pt] \scriptsize(\textit{n}=108)}}
    & \multicolumn{2}{c}{\makecell{\feedbackbotbadge \\ \textbf{FeedbackBot} \\[-2pt] \scriptsize(\textit{n}=109)}}
    & \multicolumn{2}{c}{\makecell{\genbotbadge \\ \textbf{GenBot} \\[-2pt] \scriptsize(\textit{n}=107)}} \\
    \cmidrule(lr){2-3}\cmidrule(lr){4-5}\cmidrule(lr){6-7}\cmidrule(lr){8-9}\cmidrule(lr){10-11}
    \textbf{Outcome} & \textit{M} & \textit{SD} & \textit{M} & \textit{SD} & \textit{M} & \textit{SD} & \textit{M} & \textit{SD} & \textit{M} & \textit{SD} \\
    \midrule
    \multicolumn{11}{l}{\textit{Quality related outcomes}} \\
    Goal difficulty                      & 3.75 & 0.75 & 3.55 & 0.75 & 3.40 & 0.90 & 3.71 & 0.81 & 3.44 & 0.90 \\
    Goal specificity                     & 2.15 & 1.27 & 2.74 & 1.11 & 2.63 & 1.28 & 3.41 & 0.63 & 3.62 & 0.56 \\
    Implementation Intention Quality & 0.85 & 1.20 & 2.11 & 1.05 & 1.79 & 0.78 & 3.29 & 0.74 & 3.24 & 0.71 \\
    \midrule
    \multicolumn{11}{l}{\textit{Perception related outcomes}} \\
    Goal Commitment                      & 4.22 & 0.58 & 4.06 & 0.59 & 3.93 & 0.67 & 3.96 & 0.74 & 3.96 & 0.67 \\
    Social Presence                      & 3.76 & 1.42 & 3.81 & 1.49 & 4.18 & 1.48 & 4.25 & 1.71 & 4.34 & 1.39 \\
    Intention to Act                     & 5.87 & 0.98 & 5.56 & 1.21 & 5.52 & 1.20 & 5.36 & 1.40 & 5.22 & 1.43 \\
    \midrule
    \multicolumn{11}{l}{\textit{Technology adoption related outcomes}} \\
    Effort Expectancy                    & 3.95 & 0.54 & 3.99 & 0.68 & 4.09 & 0.67 & 3.84 & 0.79 & 4.06 & 0.62 \\
    Performance Expectancy               & 3.52 & 0.84 & 3.54 & 0.88 & 3.81 & 0.86 & 3.66 & 1.03 & 3.84 & 0.74 \\
    Hedonic Motivation                   & 3.34 & 0.89 & 3.35 & 0.95 & 3.54 & 0.93 & 3.43 & 1.04 & 3.51 & 0.81 \\
    Attitude towards Chatbot                     & 3.72 & 0.71 & 3.72 & 0.81 & 3.86 & 0.76 & 3.67 & 0.94 & 3.88 & 0.72 \\
    \bottomrule
\end{tabular}
\Description{A table showing descriptive statistics for 10 study outcomes across 5 chatbot conditions. The data indicates that GenBot and FeedbackBot generally scored higher on Goal Quality and Social Presence metrics, while the ControlBot scored highest on Intention to Act.}
\end{table*}

\section{Results}\label{sec:results}
\subsection{Main Analyses of Design Features}
We first conducted omnibus Kruskal–Wallis tests to examine overall differences between conditions (see Table \ref{tab:descriptives_by_condition} for descriptive statistics). Significant effects emerged for perceived goal difficulty, $H(4)=15.08$, $p=.005$, $\eta^2=.02$, goal specificity, $H(4)=120.98$, $p<.001$, $\eta^2=.22$, and implementation intention quality, $H(4)=271.93$, $p<.001$, $\eta^2=.50$. 

\subsubsection{H1: \textbf{Guidance} alone already supported in increasing goal specificity and implementation intention quality} 

Hypothesis H1 expected a higher (a) perceived goal difficulty (b) goal specificity, and (c) quality of implementation intentions when participants receive guidance from a chatbot compared to when no guidance is provided. Regarding the perceived goal difficulty, Dunn's post hoc test revealed no significant difference between the ControlBot and the GuidanceBot ($z = -1.68, p = .719, r =-.11$). Therefore, the results do not support H1a, indicating no effect of guidance on the perceived goal difficulty. The hypotheses 1b and 1c were supported, as participants using the GuidanceBot set more specific goals (\textit{z} = 3.42, \textit{p} = .001, \textit{r} = .23) and higher quality implementation intentions (\textit{z} = 5.91, \textit{p} = <.001, \textit{$r$} = .40) compared to participants using the ControlBot. Therefore, guidance by a chatbot led to a higher goal specificity and quality of implementation intentions, but did not have a significant effect on the perceived goal difficulty (see Table \ref{tab:results_h1h3}).

\begin{table*}[t]
\caption{\textbf{Pairwise comparisons testing individual design features (H1--H3).} Reported are Dunn test $z$-statistics, Holm-adjusted one-sided $p$-values, and effect sizes ($r$). Effect sizes are calculated as $r = z/\sqrt{n_1+n_2}$. Guidance improved goal specificity and implementation intention quality compared to the control but had no effect on perceived difficulty (H1). Adaptive suggestions did not produce additional improvements beyond guidance (H2). Feedback further enhanced goal specificity and implementation intention quality beyond guidance, though not goal difficulty (H3).}
\centering
\begin{tabular}{@{}lp{4.2cm}cccc@{}} 
\toprule
Hyp. & Outcome & $z$ & $p$ & $r$ & Supported? \\
\midrule
\multicolumn{6}{l}{\textit{H1: \guidancebadge{} Guidance (GuidanceBot vs.  ControlBot)}}\\
H1a & Goal difficulty & $-1.68$ & $.719$ & $-0.11$ & \textcolor{BrickRed}{\ding{55}} \\
H1b & Goal specificity & $ 3.42$ & $.001$ & $ 0.23$ & \textcolor{ForestGreen}{\ding{51}} \\
H1c & Implementation intention quality & $ 5.91$ & $< .001$ & $ 0.40$ & \textcolor{ForestGreen}{\ding{51}} \\
\midrule
\multicolumn{6}{l}{\textit{H2: \suggestionbadge{} Suggestions (SuggestionBot vs. GuidanceBot)}}\\
H2a & Goal difficulty & $-1.22$ & $.556$ & $-0.08$ & \textcolor{BrickRed}{\ding{55}} \\
H2b & Goal specificity & $-0.02$ & $.510$ & $ 0.00$ & \textcolor{BrickRed}{\ding{55}} \\
H2c & Implementation intention quality & $-2.28$ & $.978$ & $-0.15$ & \textcolor{BrickRed}{\ding{55}} \\
\midrule
\multicolumn{6}{l}{\textit{H3: \feedbackbadge{} Feedback ( FeedbackBot vs.  GuidanceBot})}\\
H3a & Goal difficulty & $ 1.50$ & $.336$ & $ 0.10$ & \textcolor{BrickRed}{\ding{55}} \\
H3b & Goal specificity & $ 4.34$ & $< .001$ & $ 0.29$ & \textcolor{ForestGreen}{\ding{51}} \\
H3c & Implementation intention quality & $ 7.37$ & $< .001$ & $ 0.50$ & \textcolor{ForestGreen}{\ding{51}} \\
\bottomrule
\end{tabular}
\label{tab:results_h1h3}
\Description{A table showing pairwise statistical comparisons for Hypotheses 1, 2, and 3. The results indicate that H1 (Guidance) and H3 (Feedback) successfully improved Goal Specificity and Implementation Intention Quality (indicated by green checkmarks and p-values < .001). However, H2 (Suggestions) failed to produce improvements over Guidance alone (indicated by red crosses). Goal Difficulty was not significantly affected in any condition.}
\end{table*}

\subsubsection{H2: \textbf{Suggestions} did not further support goal setting}
Hypothesis H2 expected a higher (a) perceived goal difficulty (b) goal specificity and (c) quality of implementation intentions when participants received adaptive suggestions, compared to when no adaptive suggestions were provided. The results indicated no significant differences between the SuggestionBot and the GuidanceBot regarding perceived goal difficulty, \textit{z} = -1.22, \textit{p} = .556, \textit{r} = -.08, goal specificity, \textit{z} = -0.02, \textit{p} = .510, \textit{r} = .00, and quality of implementation intentions, \textit{z} = -2.28, \textit{p} = .978, \textit{r} = -.15. Thus, we did not find evidence that providing adaptive suggestions did increase perceived goal difficulty, goal specificity, or the quality of implementation intentions compared to guidance alone. 

\subsubsection{H3: Beyond guidance, \textbf{feedback} further increased the effectiveness of goal setting}
Hypothesis H3 expected a higher (a) perceived goal difficulty (b) goal specificity, and (c) quality of implementation intentions when participants receive feedback from a chatbot compared to when no feedback is provided. To test this hypothesis, we compared participants using the FeedbackBot with participants using the GuidanceBot. As with hypotheses 2, results showed no significant differences for the perceived goal difficulty (\textit{z} = 1.50, \textit{p} = .336,  \textit{r} = .10). In contrast, the differences between both groups were significant for goal specificity (\textit{z} = 4.34, \textit{p} = <.001,  \textit{r} = .29), and quality of implementation intentions (\textit{z} = 7.37, \textit{p} = <.001,  \textit{r} = .50). This indicates that feedback supports participants in setting more specific goals and higher quality implementation intentions. However, participants did not perceive the goals to be more difficult after interactively refining them with a chatbot. 

\begin{table*}[t]
\caption{\textbf{Pairwise comparisons between GenBot and other chatbot conditions.} Reported are $z$-values (with degrees of freedom), Holm-adjusted one-sided $p$-values, and effect sizes (r). \textit{GenBot} significantly improved goal specificity relative to all three single-feature chatbots, but showed no consistent advantage for goal difficulty. For implementation intention quality, significant differences were only observed against \textit{GuidanceBot} and \textit{SuggestionBot}. *p < .05, ***p < .001}
\centering
\begin{tabular}{@{}llccc@{}}
\toprule
\textbf{Outcome} & \textbf{Comparison} & \textbf{$z$ (df)} & \textbf{$p$} & \textbf{$r$} \\
\midrule
\multirow{3}{*}{Goal Difficulty} 
  & \genbotbadge{}  GenBot vs. \guidancbotbadge{} GuidanceBot   & $-1.08$ (537) & .556 & $-0.07$ \\
  & \genbotbadge{}  GenBot vs. \suggestionbotbadge{} SuggestionBot & $ 0.13$ (537) & .500 & $ 0.01$ \\
  & \genbotbadge{}  GenBot vs. \feedbackbotbadge{} FeedbackBot   & $-2.57$ (537) & .965 & $-0.18$ \\
\midrule
\multirow{3}{*}{Goal Specificity} 
  & \genbotbadge{}  GenBot vs. \guidancbotbadge{} GuidanceBot   & $ 6.35$ (538) & $<.001^{***}$ & $ 0.43$ \\
  & \genbotbadge{}  GenBot vs. \suggestionbotbadge{}SuggestionBot & $ 6.36$ (538) & $<.001^{***}$ & $ 0.43$ \\
  & \genbotbadge{}  GenBot vs. \feedbackbotbadge{} FeedbackBot   & $ 2.01$ (538) & .044$^{*}$ & $ 0.14$ \\
\midrule
\multirow{3}{*}{Implementation Intention Quality} 
  & \genbotbadge{}  GenBot vs. \guidancbotbadge{} GuidanceBot   & $ 7.00$ (538) & $<.001^{***}$ & $ 0.48$ \\
  &\genbotbadge{} GenBot vs. \suggestionbotbadge{} SuggestionBot & $ 9.26$ (538) & $<.001^{***}$ & $ 0.63$ \\
  & \genbotbadge{}  GenBot vs. \feedbackbotbadge{} FeedbackBot   & $-0.35$ (538) & .638 & $-0.02$ \\
\bottomrule
\end{tabular}
\label{tab:pairwise_comparisons}
\Description{A table showing pairwise comparisons of GenBot against GuidanceBot, SuggestionBot, and FeedbackBot. GenBot significantly outperformed all three other conditions on Goal Specificity ($p < .05$). For Implementation Intention Quality, GenBot outperformed GuidanceBot and SuggestionBot ($p < .001$) but was statistically equivalent to FeedbackBot ($p=.638$). No significant differences were found for Goal Difficulty across any comparisons.}
\end{table*}

\subsubsection{H4: Combining \textbf{all design features} did further support goal specificity}
Hypothesis H4 expected the combination of all three design features---\textit{guidance, suggestions, and feedback}---to surpass each feature alone regarding (a) goal difficulty, (b) goal specificity, and (c) quality of implementation intentions.  To test this hypothesis, we compared the GenBot with the GuidanceBot, SuggestionBot, and FeedbackBot. As with the other design features, the GenBot did not increase goal difficulty (see Table \ref{tab:pairwise_comparisons}). There was also no significant difference regarding implementation intention quality for the GenBot compared to the FeedbackBot (\textit{z} = -.35, \textit{p} = .638,  \textit{r} = -.02).
However, participants interacting with the GenBot showed significantly higher goal specificity compared to participants interacting with the GuidanceBot  (\textit{z} = 6.35, \textit{p} = < .001,  \textit{r} = .43), and the SuggestionBot  (\textit{z} = 6.36, \textit{p} = < .001,  \textit{r} = .43). The contrast with the FeedbackBot also reached significance (\textit{z} = 2.01, \textit{p} = .044, \textit{r} = .14), though this effect is only marginal and should be interpreted with caution given the overall number of tests \textsuperscript{(2AC)}. 
In summary, while the GenBot did not increase goal difficulty or implementation intention quality beyond single features, it significantly increased goal specificity, indicating that combining all three mechanisms yields the highest effectiveness for goal setting.

\subsubsection{H5: GenBot 
fostered \textbf{goal commitment} via \textbf{social presence}, but reduced it overall}
\begin{figure}[tbp]
    \centering
    \includegraphics[width=0.45\textwidth]{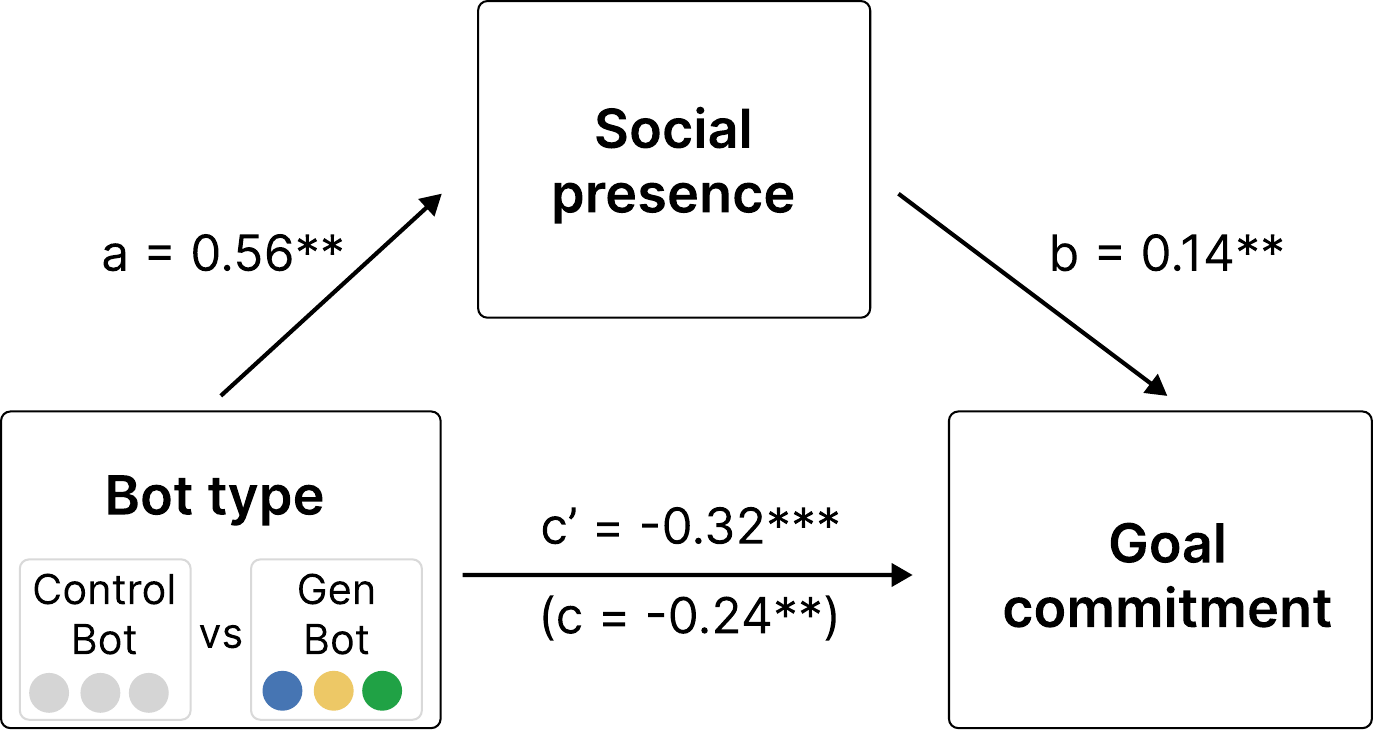}
    \caption{\textbf{Mediation model showing that social presence mediates the relationship between chatbot type and goal commitment.} Compared to the \textit{ControlBot}, the \textit{GenBot} increased perceived social presence, which in turn predicted higher goal commitment. The direct effect of chatbot type on goal commitment remained negative, indicating a suppression effect.}
    \label{fig:mediation}
\Description{This mediation model illustrates that using the GenBot (versus ControlBot) significantly increases perceived social presence (a = 0.56, p < .01), which then positively influences goal commitment (b = 0.14, p < .01). The direct effect of chatbot type on goal commitment is negative (c′ = –0.32, p < .001), with a total effect of c = –0.24, p < .01. Thus, the figure conveys that social presence partly mediates the relationship between chatbot design and user commitment.}
\end{figure}

Hypothesis H5 expected social presence to mediate the relationship between chatbot type (GenBot vs. ControlBot) and goal commitment. Perceived social presence was significantly higher when participants interacted with the GenBot compared to the ControlBot ($t(216) = 2.92$, $p = .004$, $b^{*} = 0.52$, $SE = .20$, 90\% CI $[.19, .84]$). Also, there was a positive correlation between perceived social presence and participants’ goal commitment ($t(215) = 3.65$, $p < .001$, $b^{*} = 0.11$, $SE = .03$, 90\% CI $[.06, .15]$), indicating that individuals who perceived higher social presence tend to exhibit greater goal commitment (see Figure~\ref{fig:mediation}). Additionally, the indirect relationship between the type of chatbot and participants’ goal commitment was significantly mediated by perceived social presence (90\% CI $[.03, .14]$). However, the total effect of the type of chatbot on goal commitment was not significant and actually negative ($t(216) = -1.14$, $p = .127$, $b^{*} = -0.10$, $SE = .09$, 90\% CI $[-.24, .04]$), indicating that participants interacting with the GenBot exhibited lower goal commitment. Detailed results are in Appendix~\ref{appendix:mediation}.

In summary, the GenBot has two opposing effects on goal commitment, namely a positive indirect effect through increased social presence and an even stronger negative effect. Indications of these results will be discussed in Section~\ref{sec:discussion}.

\subsection{Exploratory Analysis}
\subsubsection{Social Presence Mediation Generalizes to All LLM-Based Conditions}

\new{To examine whether social presence functions as a general mediating mechanism across different chatbot conditions, we conducted two complementary post-hoc analyses. First, a multicategorical mediation analysis with the ControlBot as the reference group revealed that social presence significantly mediated the effects for all LLM-based conditions, namely, the SuggestionBot (90\% CI $[.02, .11]$), FeedbackBot (90\% CI $[.02, .12]$) and GenBot (90\% CI $[.03, .13]$). The indirect effect for GuidanceBot did not reach significance (90\% CI $[-.04, .05]$).}

\new{Second, we conducted a linear mediation analysis with the chatbot conditions reflecting increasing sophistication. Model comparisons indicated adequate fit of linear specifications for all mediation paths ($p > .154$ for all tests). This analysis showed that greater sophistication was associated with a positive indirect effect on goal commitment through social presence (effect $0.02$ (90\% CI $[0.01, 0.03]$). Across both analyses, the suppression effect identified in the main model remained. Detailed results are provided in Appendix~\ref{appendix:mediation}. Overall, these findings suggest that the mediation effect observed for Hypothesis 5 generalizes across LLM-based conditions, and that more sophisticated chatbot designs may strengthen this mediation effect.\textsuperscript{(2AC)}}

\subsubsection{Higher intention to act without any design features}
As specified in our preregistration for exploratory purposes, we asked the participants in an item about their intention to act. Interestingly, the intention to act was significantly higher for participants interacting with the ControlBot compared to participants interacting with the GenBot (\textit{p} = .005,  \textit{r} = .24). This indicates that collaboratively formulating the goals with an LLM-based chatbot might decrease the actual intention to act on the goal. 

\subsubsection{No differences regarding goal attainability}
As preregistered, we conducted an exploratory analysis examining participants’ perceived goal attainability. Group differences between the five LLM-based chatbot conditions were assessed using a Kruskal–Wallis test. Results showed no significant differences between the five conditions (\textit{H}(4) = 3.68, \textit{p} = .452). 

\subsubsection{No differences regarding technology adoption}
As part of the preregistered exploratory analysis, we included questionnaires on technology adoption. To examine possible differences between groups, we conducted a separate Kruskal-Wallis test for each subscale (see Appendix \ref{technology_adoption}). There was no difference in the subscales regarding hedonic motivation (\textit{H}(4) = 4.71, \textit{p} = .318) nor attitude towards the chatbot (\textit{H}(4) = 5.64, \textit{p} = .227). The Kruskal-Wallis test for effort expectancy yielded a significant result (\textit{H}(4) = 11.30, \textit{p} = .023), but none of the comparisons via Dunn's test were significant. There was only a significant difference between the ControlBot and the SuggestionBot regarding performance expectancy (\textit{p} = <.007, \textit{$\eta^2$} = .18). Therefore, the participants perceived the technology adoption mostly the same regardless of the chatbot they interacted with.

\section{Discussion}
\label{sec:discussion}
\subsection{Summary of Key Findings}
To our knowledge, this is the first large-scale study to disentangle the individual and combined effects of guidance, feedback, and suggestions in LLM-based goal setting chatbots. Our factorial design provides novel evidence that conversational guidance and feedback reliably improved the quality of participants’ goals and implementation intentions, whereas suggestions played a more limited role. Strikingly, more sophisticated chatbots reduced commitment and intention to act, despite producing technically higher-quality outputs. This quality–motivation trade-off offers both theoretical insight and practical design guidance, with broad relevance for behavioral domains where effective goal setting is critical, including health, education, and workplace productivity.

\subsection{Implications for Psychological Theory on Goal Setting and Implementation Intentions}

Our findings extend goal setting theory by showing that goal \emph{specificity}—long identified as a core determinant of effectiveness \citep{locke_building_2002}—can be reliably supported in a conversational LLM interface. We also observe strong improvements in the quality of implementation intentions, indicating that LLMs provide a scalable way to help people formulate effective if–then plans. In doing so, our intervention addresses a central challenge in self-regulation: assisting users in articulating clear, high-quality goals and plans. Methodologically, our approach offers a scalable experimental lever for researchers to systematically manipulate goal specificity and implementation intention quality. Thereby, we aim to advance both goal setting theory and implementation intention research.

Contrary to our expectations, LLM-based support through \textit{guidance}, \textit{suggestions}, or \textit{feedback} did not affect participants' perceived goal difficulty. Although psychological theory implies that effective goal setting comes with a relatively high goal difficulty, our findings could be explained by methodological artifacts. In our study, we measured perceived goal difficulty and did not analyze the difficulty more objectively by external raters. Moreover, it is possible that for reasons of impression management (cf.~\cite{gibson2010}), when asked to formulate goals within an experiment as well as by another entity (here: the chatbot), participants might have formulated rather difficult goals, leading to a ceiling effect, which potentially restricted variance and obscured effects.

Our results also reveal an important tension between quality and motivation. While our analyses indicate that LLM-based chatbots can increase goal commitment indirectly through heightened perceived social presence, this positive pathway was offset by stronger counteracting effects. As a result, both goal commitment and intention were even higher in the control group. 
\new{ This pattern, while counterintuitive to goal setting theory, may be explained by the study design itself. Participants in our study could freely choose goals without constraints or support in selecting appropriate goals in the first place. During the process of making goals more specific through guidance and feedback, participants may have gained deeper insight into what pursuing these goals would actually entail. A greater awareness could have triggered several demotivating mechanisms: it may have raised cost salience, making the effort and sacrifices required more salient~\cite{Wu2023}, reduced perceived volition~\cite{alberts2024} and confidence~\cite{Albers2023}  or triggered psychological reactance~\cite{Zhang2011}. Construal level theory, for example, suggests that highly concrete, low-level construals can emphasize effort and difficulty, whereas more abstract framings may sustain motivational energy \cite{trope2010}.
An alternative explanation centers on autonomy: the more elaborate chatbot may have inadvertently reduced user autonomy, thereby decreasing commitment. Self-determination theory emphasizes that preserving autonomy is critical for sustained goal pursuit~\citep{ryan2000}. This interpretation aligns with recent findings showing that AI-generated "optimal" goals can undermine motivation when they feel imposed rather than self-chosen~\citep{Lieder2022}. 
Together, these mechanisms point to a quality-motivation trade-off: while more specific goals enhance clarity, when perceived as imposed or not entirely intrinsically motivated, they may simultaneously reduce commitment and  immediate intention to act.
However, reduced commitment might not be necessarily negative; it may instead be based on a more accurate assessment of people's abilities and the effort required. In this sense, lower commitment scores could serve a protective function, preventing users from sticking to goals that, once fully specified, seem less suitable for their true aims.
Since these findings may be attributable to our chatbots implementation, the following section discusses how they can inform design decisions and proposes mediation strategies to preserve user motivation in LLM-based goal-setting interventions.
\textsuperscript{(2AC,R2)}
}

\subsection{Chatbot Mechanisms and Design Recommendations}
\new{Our findings demonstrate that LLM-based chatbots offer great potential for generating high-quality goals and implementation intentions, though they also might introduce drawbacks for certain domains.}
The results indicate distinct contributions of each chatbot feature on goal setting and implementation intention outcomes, highlighting potential recommendations for future HCI research and practitioners working on behavior change interventions.

\textit{\textbf{\new{Use guidance---as an effective mechanism without the need for LLMs.\textsuperscript{(2AC)}}}}
Providing structured conversational scaffolds improved goal specificity and implementation intention quality. This finding resonates with earlier HCI work showing that guidance alone can already be effective without the need for LLMs \citep{lim_chatbot-delivered_2022,abd-alrazaq_effectiveness_2020}. Guidance thus remains the most reliable building block for digital goal setting.

\textit{\textbf{\new{ Use feedback---to help users apply well-defined frameworks.\textsuperscript{(2AC)}}}}
Feedback produced the strongest quality gains. In our study on goal setting and implementation intention formation, the desired output was clearly specified, allowing required elements to be checked against the input and enabling actionable, context-appropriate LLM feedback. For domains in which quality criteria can be operationalized as checklists or structured templates, designers should explicitly define intervention stages and optimal outcomes to support effective LLM-based feedback.

\textit{\textbf{\new{ Use Suggestions---as complement to feedback rather than a standalone driver.\textsuperscript{(2AC)}}} }
Contrasting other prior studies where adaptive and generative suggestions effectively fostered creativity and novel idea generation \citep{doshi_generative_2024, peng_impact_2023}, standalone suggestions did not significantly enhance quality and appear most useful as prompts for implementing feedback rather than as substitutes for user-authored content. When combined, suggestions helped users act on feedback more efficiently, making revisions easier to implement. Chatbots should therefore intertwine guidance, feedback, and light-touch suggestions rather than deploying them in isolation.

\new{\textit{\textbf{ Protect commitment---by helping users discover an appropriate goal first.}} In our study, participants were asked to name a single goal without receiving any prior support in choosing an appropriate goal in the first place. To preserve commitment while formulating a goal and translating it into implementation intentions, an AI intervention could first guide users to reflect on goal hierarchies to identify superordinate goals that truly resonate with them \citep{carver_control_1982, hochli_focusing_2018}. Furthermore, techniques such as Mental Contrasting~\cite{oettingen_expectancy_2000} — the practice of comparing a desired future against present reality— could reduce the chance that users only realize during the specification process that they were not genuinely motivated by their initial goal, as it forces a realistic evaluation of obstacles before commitment is made \citep{dekker_optimizing_2020}.\textsuperscript{(2AC)}}

\textit{\textbf{\new{ Protect commitment---by preserving user autonomy.}}}
\new{Considering the lower goal commitment and intention to act in individuals who interacted with more elaborate LLM chatbots\textsuperscript{(2AC)}}, interventions must preserve user autonomy and volition. Autonomy-supportive architecture \citep{ryan2000}, reaffirming superordinate goals \citep{hochli2018}, and explicitly offering user flexibility may counteract motivational costs when concrete implementation steps are introduced. \new{ For instance, in recreational or exploratory domains---such as developing a new hobby or social pursuits---overly structured goal specification may stifle intrinsic motivation and the open-ended discovery that sustains long-term engagement \citep{etkin_hidden_2016, ordonez_goals_2009}.\textsuperscript{(2AC)}} Accordingly, LLM-based interventions should adopt a supportive human-led role instead of outsourcing the primary mental work to the LLM \cite{maier_partnering_2025}. This appears particularly relevant as technologies that can support users in goal setting and beyond become increasingly autonomous and can proactively intervene in user routines, thus potentially restricting perceived user autonomy. 

\textit{\textbf{\new{Elicit commitment---by using social presence.\textsuperscript{(2AC)}}}}  
The \textit{social presence} elicited by the LLM-based chatbot was significantly stronger, supporting existing research suggesting that more interactive and humanized LLM interfaces foster social connections \cite{go_humanizing_2019}. We observed that social presence mediated higher goal commitment. Chatbots that maintain goals at a less specific level, and use LLMs to act more as accountability partners--thereby \new{leveraging principles of \textit{supportive accountability} \citep{mohr_model_2011}}---might therefore potentially raise goal commitment.


\subsection{Limitations and Future Research}
As with other studies, ours is not free of limitations, which open interesting possibilities for future research.
\new{First, our study examined the immediate effects of LLM-specific design features on goal formulation rather than on goal pursuit and attainment. Future research should investigate whether the quality improvements observed through these features also translate to better goal attainment over time.\textsuperscript{(2AC)}}
Second, the uniformly high ratings for goal commitment and perceived difficulty across conditions suggest possible ceiling effects or social desirability biases. More discriminating measures or behavioral indicators might provide greater sensitivity to differences between conditions. Third, on a more abstract level, our study examined how LLM-based chatbots can refine users’ stated goals, but it did not address the challenge of identifying which goals to pursue in the first place. For instance, while one user may initially aspire to run a marathon, a different goal might ultimately be more realistic, beneficial, or aligned with their circumstances. To provide truly effective support, future research should investigate how chatbots can assist not only in refining goal formulation but also in guiding users toward selecting meaningful and appropriate goals. \new{Fourth, although we explored several iterations of the three features during the prototyping phase, the final implementation reflects only one possible design. Different versions of the features might have yielded different effects, potentially producing stronger outcomes for elements such as the suggestion feature.\textsuperscript{(2AC)}} \new{Fifth, our design involved multiple outcomes and hypotheses. To account for family-wise error rate, we Holm-adjusted $p$-values within each set of hypotheses. Nevertheless, given the total number of analyses, isolated findings with small effect sizes and $p$-values close to the significance threshold (the small GenBot–FeedbackBot difference in goal specificity) should be interpreted with caution. \textsuperscript{(2AC)}}

\section{Conclusion}
\label{sec:conclusion}
Many people face challenges in acting on their goals. However, it is unclear if and how LLM-based chatbots can help people set more effective goals and actions. Our study demonstrates that LLM assistance can help people set goals and actions more effectively by helping them adopt goal setting theory and implementation intentions. We show how different LLM-based chatbot features contribute distinct benefits but also show some trade-offs. These findings extend goal setting theory and implementation intentions and contribute to HCI research by emphasizing the importance of not only focusing on the outcomes of LLM-based interventions but also examining the foundational building blocks.

\begin{acks}
We would like to thank Julia Falinska for her valuable support in designing the graphics; Roman Mayr and Robin Frasch for helping to test the goal-setting application; and Emilia de Mattia for helping with the rating of the goals and implementation intentions.
\end{acks}

\bibliographystyle{ACM-Reference-Format}
\bibliography{software.bib}

\newpage

\appendix
\onecolumn
\section{Rating Criteria for goal specificity and quality of implementation intentions}
\begin{table}[htpb]
    \centering
    \caption{Rating criteria for the goal specificity based on the two categories "when" and "where"}
    \label{tab:goal_rating}
    \scriptsize
    \begin{tabularx}{\textwidth}{|c|c|c|X|X|X|X|}
        \hline
        \textbf{Rating Categories} & \textbf{Scale} & \textbf{Class} & \textbf{Description} & \textbf{Example: Running} & \textbf{Example: Anxiety} & \textbf{Example: PhD} \\
        \hline
        When & 0 & Not Provided & No time frame is mentioned for when the goal will be achieved. & Run 10k & Reduce anxiety & Apply for PhD programs \\
        \hline
        & 1 & General & A time frame is mentioned, but it is vague or broad. & Soon & As soon as possible & Sometime next year \\
        \hline
        & 2 & Specific & A clear and specific time frame is provided. Example: \textit{"By December 1st."} & By December 1st. & By November 1st & Submit PhD applications by January 15th \\
        \hline
        Measurement & 0 & Not Provided & No measurable outcome is given to assess the success of the goal. & Improve running & Feel less anxious & Find PhD opportunities \\
        \hline
        & 1 & General & The goal has a vague or broad measurement. & I want to complete the marathon & I want to feel less anxious in social situations & I want to be accepted into a PhD program \\
        \hline
        & 2 & Specific & A concrete and measurable outcome is provided. Example: \textit{"Run 5 km in under 20 minutes"} & Run 5k in under 20 minutes & Reduce anxiety score by 20\% in three months, as measured by the GAD-7 scale & Submit 5 applications by January and get accepted into at least one program by June \\
        \hline
    \end{tabularx}

\Description{A coding scheme table defining a 3-point rating scale (0 to 2) for Goal Specificity. It assesses two dimensions: 'When' (time frame) and 'Measurement' (outcome). For both dimensions, a score of 0 indicates 'Not Provided', 1 indicates 'General/Vague', and 2 indicates 'Specific'. The table includes examples for running, anxiety, and PhD applications for each level of the scale.}
\end{table}

\begin{table}[htpb]
    \centering
    \caption{Rating criteria for the quality of implementation intentions based on the categories "General Format", "Appropriate Action", "Clear Time", and "Place".}
    \label{tab:goal_categorization}
    \scriptsize
    \begin{tabularx}{\textwidth}{|c|c|c|X|X|X|X|}
        \hline
        \textbf{Rating Categories} & \textbf{Scale} & \textbf{Class} & \textbf{Description} & \textbf{Example: Running} & \textbf{Example: Anxiety} & \textbf{Example: PhD} \\
        \hline
        General Format & 0 & Not Provided & If rated as 0, no more rating should be conducted and the total rating is 0 & I will run tomorrow at 6pm & I want to feel better & If I think about my research every day for 5 hours, then I will complete a new paper in a few months. \\
        \hline
        & 1 & Provided & An action is connected with a cue in an if-then statement (or similar, e.g., when-then) &  &  &  \\
        \hline
        Appropriate Action & 0 & Not Provided & The action is not relevant or suited to achieving the stated goal type (learning or performance). & Goal: Learn about 5 strategies to train for the marathon. Action: When I come home from work, I go for a run & Goal: Learn 5 strategies to manage anxiety effectively. Action: "When I come home from work, I will research one anxiety management technique and write it down in my journal." & Goal: Publish three papers until January 1st, 2026; Action: When I get up, I'll learn about research methods \\
        \hline
        & 1 & Provided & The action is relevant and logically suited to achieving the stated goal type (learning or performance). & Action: When I come home from work, I'll research training strategies & Action: "If I start to feel overwhelmed or anxious, then I will immediately practice the 5-4-3-2-1 grounding technique" & Goal: Publish three papers until January 1st, 2026; Action: When I get up, I'll work on the method for my study \\
        \hline
        Clear Time & 0 & Not Provided & No specific day or time is mentioned when the activity will take place. & "Run sometime" & Practice relaxation techniques regularly & Work on PhD whenever possible \\
        \hline
        & 1 & Provided & A specific day or time is mentioned for performing the activity. & When I come home on Monday and Thursday & Practice relaxation techniques whenever I feel stressed & Study every Tuesday and Thursday \\
        \hline
        Place & 0 & Not Provided & No location or place is provided for the activity. & "Go running" & "Relax" & "Work on research" \\
        \hline
        & 1 & Provided & A clear location is provided for where the activity will take place. & "Run at the park" & "Practice mindfulness at home" & "Study in the university library" \\
        \hline
    \end{tabularx}
\Description{A coding scheme table defining the rating criteria for Implementation Intention Quality based on four binary categories (scored 0 or 1). The categories are: 1) General Format (requires an 'if-then' structure; a score of 0 here results in a total score of 0); 2) Appropriate Action (relevance to goal type); 3) Clear Time; and 4) Place. The table provides descriptions and examples (Running, Anxiety, PhD) for both the 'Not Provided' (0) and 'Provided' (1) classifications.}
\end{table}
\newpage

\FloatBarrier
\section{Manipulation Check}

\begin{table}[htbp]
\centering
\caption{Multiple linear regression (OLS) results for chatbot group effects on guidance perception, effort expectancy, and intention to act. Unstandardized coefficients ($B$), standard errors (SE), $t$-values, and $p$-values are reported. Significance stars indicate conventional thresholds.}
\label{tab:mlr_results}
\begin{tabular}{@{}llcccc@{}}
\toprule
\textbf{Outcome} & \textbf{Predictor} & $B$ & SE & $t$ & $p$ \\
\midrule
\multirow{5}{*}{Guidance} 
  & Intercept         & 3.39 & 0.09 & 39.53 & $<.001^{***}$ \\
  & GuidanceBot       & 0.40 & 0.12 & 3.28  & .001$^{**}$ \\
  & SuggestionBot     & 0.67 & 0.12 & 5.47  & $<.001^{***}$ \\
  & FeedbackBot       & 0.48 & 0.12 & 3.96  & $<.001^{***}$ \\
  & GenBot            & 0.87 & 0.12 & 7.15  & $<.001^{***}$ \\
\midrule
\multirow{5}{*}{Suggestions} 
  & Intercept         & 4.12 & 0.10 & 43.00 & $<.001^{***}$ \\
  & ControlBot        & -0.83 & 0.13 & -6.18 & $<.001^{***}$ \\
  & GuidanceBot       & -0.98 & 0.14 & -7.24 & $<.001^{***}$ \\
  & FeedbackBot       & -0.55 & 0.14 & -4.03 & $<.001^{***}$ \\
  & GenBot            & 0.05 & 0.14 & 0.35  & .725 \\
\midrule
\multirow{5}{*}{Feedback} 
  & Intercept         & 3.85 & 0.09 & 40.74 & $<.001^{***}$ \\
  & ControlBot        & -0.74 & 0.13 & -5.60 & $<.001^{***}$ \\
  & SuggestionBot     & -0.03 & 0.13 & -0.21 & .836 \\    
  & GuidanceBot       & -0.51 & 0.13 & -3.84 & $<.001^{***}$ \\
  & GenBot            & 0.32 & 0.13 & 2.36  & .019$^{*}$ \\
\bottomrule
\end{tabular}

\vspace{4pt}
\raggedright
\textit{Note}. Reference groups differ per regression: For \textbf{Guidance}, \textit{ControlBot} is the reference. For \textbf{Suggestions}, \textit{SuggestionBot} is the reference. For \textbf{Feedback}, \textit{FeedbackBot} is the reference. $^{*}p<.05$, $^{**}p<.01$, $^{***}p<.001$.
\Description{A table reporting OLS regression results for three outcomes: Guidance, Suggestions, and Feedback. For Guidance (ref: ControlBot), all chatbot conditions showed significant positive effects, with GenBot showing the strongest coefficient (B=0.87). For Suggestions (ref: SuggestionBot), GenBot was statistically equivalent (p=.725), while others were significantly lower. For Feedback (ref: FeedbackBot), GenBot scored significantly higher (B=0.32), while Control and Guidance scored significantly lower.}
\end{table}

\FloatBarrier
\section{Technology Adoption}\label{technology_adoption}

\begin{table}[htbp]
  \caption{Kruskal–Wallis Test Results for Chatbot Evaluation Measures Across Groups}
  \centering
  \begin{tabular}{lcccc}
    \toprule
    \textbf{Measure} & \textbf{\emph{H}(4)} & \textbf{\emph{p}} & \textbf{$\eta^2_H$} & \textbf{Sig.} \\
    \midrule
    Effort Expectancy       & 11.30 & .023 & .014  & * \\
    Performance Expectancy  & 13.09 & .011 & .017  & * \\
    Hedonic Motivation      & 4.71  & .318 & .001  &   \\
    Attitude toward Chatbot & 5.64  & .227 & .003  &   \\
    Goal Attainability      & 3.68  & .452 & <.001 &   \\
    \bottomrule
  \end{tabular}
  \label{tab:kruskal-wallis-results}
\Description{A table reporting Kruskal–Wallis test results for five chatbot evaluation measures. Statistically significant differences across groups were found for Effort Expectancy (p=.023) and Performance Expectancy (p=.011). No significant differences were observed for Hedonic Motivation, Attitude toward Chatbot, or Goal Attainability}
\end{table}

\FloatBarrier
\newpage
\section{Mediation Analysis}\label{appendix:mediation}

\begin{table}[htbp]
\centering
\caption{Mediation Analysis of the Effect of Chatbot Condition (GenBot vs ControlBot) on Goal Commitment through Social Presence}
\label{tab:mediation}
\begin{tabular}{@{}lccccc@{}}
\toprule
\textbf{Path/Effect} & \textbf{Coefficient} & \textbf{SE} & \textbf{\textit{t}} & \textbf{\textit{p}} & \textbf{95\% CI} \\
\midrule
\multicolumn{6}{@{}l@{}}{\textit{Path a}: Chatbot Condition $\rightarrow$ Social Presence} \\
\quad Constant & 3.75 & 0.13 & 28.00 & $< .001$ & [3.53, 3.98] \\
\quad Group (GenBot vs. ControlBot) & 0.56 & 0.19 & 2.92 & .004 & [0.24, 0.87] \\
\quad Model $R^2$ & \multicolumn{5}{@{}l@{}}{.038, \textit{F}(1, 216) = 8.52, \textit{p} = .004} \\
\midrule
\multicolumn{6}{@{}l@{}}{\textit{Path b \& c'}: Social Presence and Chatbot Condition $\rightarrow$ Goal Commitment} \\
\quad Constant & 3.66 & 0.12 & 30.01 & $< .001$ & [3.46, 3.86] \\
\quad Group (GenBot vs. ControlBot) & -0.32 & 0.08 & -3.94 & $< .001$ & [-0.46, -0.19] \\
\quad Social Presence & 0.14 & 0.03 & 5.03 & $< .001$ & [0.10, 0.19] \\
\quad Model $R^2$ & \multicolumn{5}{@{}l@{}}{.138, \textit{F}(2, 215) = 17.22, \textit{p} < .001} \\
\midrule
\multicolumn{6}{@{}l@{}}{\textit{Path c}: Total Effect of Chatbot Condition $\rightarrow$ Goal Commitment} \\
\quad Constant & 4.20 & 0.06 & 70.31 & $< .001$ & [4.10, 4.30] \\
\quad Group (GenBot vs. ControlBot) & -0.24 & 0.09 & -2.86 & .005 & [-0.38, -0.10] \\
\quad Model $R^2$ & \multicolumn{5}{@{}l@{}}{.037, \textit{F}(1, 216) = 8.18, \textit{p} = .005} \\
\midrule
\multicolumn{6}{@{}l@{}}{\textbf{Mediation Effects}} \\
\midrule
\textbf{Effect} & \textbf{Estimate} & \textbf{Boot SE} & & & \textbf{90\% Boot CI} \\
\midrule
Total Effect (\textit{c}) & -0.24 & 0.09 & & .005 & [-0.38, -0.10] \\
Direct Effect (\textit{c'}) & -0.32 & 0.08 & & $< .001$ & [-0.46, -0.19] \\
Indirect Effect (\textit{ab}) & 0.08 & 0.03 & & & [0.03, 0.14] \\
\quad Partially Standardized & 0.13 & 0.05 & & & [0.05, 0.21] \\
\bottomrule
\end{tabular}

\par\vspace{1ex}
\parbox{\linewidth}{
\small
\textit{Note.} \textit{N} = 218. Chatbot condition coded as 0 = ControlBot, 1 = GenBot. Bootstrap confidence intervals based on 20,000 bootstrap samples. The indirect effect is significant as the 90\% confidence interval does not contain zero, indicating significant mediation through social presence.
}
\Description{A mediation analysis table showing that GenBot significantly increased Social Presence (Path a, B=0.56), which in turn positively predicted Goal Commitment (Path b, B=0.14). The indirect effect was significant and positive (B=0.08, 90\% CI [0.03, 0.14]). However, the direct effect of GenBot on Goal Commitment remained negative (B=-0.32), suggesting that Social Presence partially mitigated an otherwise negative relationship.}
\end{table}


\begin{table}[ht]
\color{black}
\caption{\new{\textbf{Mediation analysis: Social Presence mediating the effect of condition on Goal Commitment.} Reported are unstandardized coefficients with 90\% confidence intervals. Indirect effects are based on 20,000 bootstrap samples. The reference category is the ControlBot. $^\dagger p < .10$, *$p < .05$, **$p < .01$, ***$p < .001$}}
\centering
\begin{tabular}{@{}llccccc@{}}
\toprule
\textbf{Path} & \textbf{Comparison} & \textbf{$b$} & \textbf{SE} & \textbf{$t$} & \textbf{$p$} & \textbf{90\% CI} \\
\midrule
\multicolumn{7}{l}{\textit{Effects on Social Presence (a-paths)}} \\
  & \guidancbotbadge{} GuidanceBot vs. \controlbotbadge{} ControlBot & $0.04$ & $0.20$ & $0.20$ & $.841$ & $[-0.29, 0.37]$ \\
  & \suggestionbotbadge{} SuggestionBot vs. \controlbotbadge{} ControlBot & $0.44$ & $0.20$ & $2.19$ & $.029^{*}$ & $[0.11, 0.78]$ \\
  & \feedbackbotbadge{} FeedbackBot vs. \controlbotbadge{} ControlBot & $0.49$ & $0.20$ & $2.42$ & $.016^{*}$ & $[0.16, 0.83]$ \\
  & \genbotbadge{} GenBot vs. \controlbotbadge{} ControlBot & $0.56$ & $0.20$ & $2.75$ & $.006^{**}$ & $[0.22, 0.89]$ \\
\midrule
\multicolumn{7}{l}{\textit{Effect of Social Presence on Goal Commitment (b-path)}} \\
  & Social Presence $\rightarrow$ Goal Commitment & $0.14$ & $0.02$ & $7.75$ & $<.001^{***}$ & $[0.11, 0.17]$ \\
\midrule
\multicolumn{7}{l}{\textit{Direct effects on Goal Commitment (c$'$-paths)}} \\
  & \guidancbotbadge{} GuidanceBot vs. \controlbotbadge{} ControlBot & $-0.15$ & $0.08$ & $-1.82$ & $.070^{\dagger}$ & $[-0.29, -0.01]$ \\
  & \suggestionbotbadge{} SuggestionBot vs. \controlbotbadge{} ControlBot & $-0.33$ & $0.08$ & $-3.89$ & $<.001^{***}$ & $[-0.47, -0.19]$ \\
  & \feedbackbotbadge{} FeedbackBot vs. \controlbotbadge{} ControlBot & $-0.31$ & $0.08$ & $-3.63$ & $<.001^{***}$ & $[-0.44, -0.17]$ \\
  & \genbotbadge{} GenBot vs. \controlbotbadge{} ControlBot & $-0.32$ & $0.08$ & $-3.79$ & $<.001^{***}$ & $[-0.46, -0.18]$ \\
\multicolumn{7}{l}{\textit{Indirect effects via Social Presence}} \\
  & \guidancbotbadge{} GuidanceBot vs. \controlbotbadge{} ControlBot & $0.01$ & $0.03$ & --- & --- & $[-0.04, 0.05]$ \\
  & \suggestionbotbadge{} SuggestionBot vs. \controlbotbadge{} ControlBot & $0.06$ & $0.03$ & --- & --- & $[0.02, 0.11]^{*}$ \\
  & \feedbackbotbadge{} FeedbackBot vs. \controlbotbadge{} ControlBot & $0.07$ & $0.03$ & --- & --- & $[0.02, 0.12]^{*}$ \\
  & \genbotbadge{} GenBot vs. \controlbotbadge{} ControlBot & $0.08$ & $0.03$ & --- & --- & $[0.03, 0.13]^{*}$ \\
\bottomrule
\end{tabular}
\label{tab:mediation_social_presence}
\Description{A mediation analysis table showing how Social Presence mediates the link between Chatbot Conditions and Goal Commitment (reference: ControlBot). Results indicate significant positive indirect effects for SuggestionBot, FeedbackBot, and GenBot (but not GuidanceBot), meaning these chatbots increased Social Presence which boosted Commitment. However, the direct effects of these chatbots on Commitment were significantly negative (approx. b = -0.32), suggesting Social Presence partially mitigated an otherwise negative impact.}
\end{table}


\begin{table}[htbp]
  \color{black}
  \centering
  \caption{\new{\textbf{Linearity Tests for Mediation Model Specification.} To validate the linear treatment of chatbot condition, we compared linear specifications against categorical alternatives for paths involving condition, and tested for non-linearity in the relationship between social presence and goal commitment. Non-significant \textit{p}-values across all paths support the linear specification used in our mediation analysis.}}
  \label{tab:linearity_tests}
  
  \begin{tabular}{@{}llccc@{}}
    \toprule
    \textbf{Path} & \textbf{Test} & \textbf{\textit{F}} & \textbf{df} & \textbf{\textit{p}} \\
    \midrule
    Path a: Condition → Social Presence           & Linear vs. Categorical & 0.57 & 3, 538 & .635 \\
    Path c: Condition → Goal Commitment (total)   & Linear vs. Categorical & 1.25 & 3, 538 & .290 \\
    Path c': Condition → Goal Commitment (direct) & Linear vs. Categorical & 1.76 & 3, 537 & .154 \\
    Path b: Social Presence → Goal Commitment     & Linear vs. Quadratic   & 0.01 & 1, 539 & .907 \\
    \bottomrule
    
    \multicolumn{5}{p{\linewidth}}{\small \textit{Note.} \textit{N} = 543. All \textit{p} > .05 indicate adequate fit of linear specifications. Path b quadratic coefficient: $\beta = -0.001$, SE = 0.011. For paths involving chatbot condition, categorical specifications included four dummy codes with Condition 0 (ControlBot) as reference.}
  \end{tabular}
\Description{A table reporting validation tests for the mediation model's linearity assumptions. It compares linear specifications against categorical or quadratic alternatives for paths a, b, c, and c'. All tests resulted in non-significant p-values (p > .05), confirming that the linear model specification provides an adequate fit for the data.}
\end{table}

\begin{table}[htbp]
\color{black}
\centering
\caption{\new{\textbf{Linear mediation analysis: Social Presence mediating the effect of chatbot sophistication on Goal Commitment.} Chatbot condition is treated as an ordinal predictor (0--4) reflecting increasing AI sophistication. Reported are unstandardized coefficients with 90\% confidence intervals. Indirect effects are based on 20,000 bootstrap samples. Linear specifications demonstrated adequate fit across all mediation paths ($p > .154$ for all tests).}}
\label{tab:mediation_extended}

\begin{tabular}{@{}lccccc@{}}
\toprule
\textbf{Path/Effect} & \textbf{Coefficient} & \textbf{SE} & \textbf{\textit{t}} & \textbf{\textit{p}} & \textbf{90\% CI} \\
\midrule
\multicolumn{6}{@{}l@{}}{\textit{Path a}: Chatbot Condition $\rightarrow$ Social Presence} \\
\quad Constant & 3.75 & 0.11 & 33.85 & $< .001$ & [3.57, 3.93] \\
\quad Chatbot Condition & 0.16 & 0.05 & 3.45 & $< .001$ & [0.08, 0.23] \\
\quad Model $R^2$ & \multicolumn{5}{@{}l@{}}{.022, \textit{F}(1, 541) = 11.94, \textit{p} $< .001$} \\
\midrule
\multicolumn{6}{@{}l@{}}{\textit{Paths b \& c'}: Social Presence and Chatbot Condition $\rightarrow$ Goal Commitment} \\
\quad Constant & 3.63 & 0.08 & 44.76 & $< .001$ & [3.49, 3.76] \\
\quad Chatbot Condition (\textit{c'}) & -0.08 & 0.02 & -4.19 & $< .001$ & [-0.11, -0.05] \\
\quad Social Presence (\textit{b}) & 0.14 & 0.02 & 7.66 & $< .001$ & [0.11, 0.17] \\
\quad Model $R^2$ & \multicolumn{5}{@{}l@{}}{.112, \textit{F}(2, 540) = 34.16, \textit{p} $< .001$} \\
\midrule
\multicolumn{6}{@{}l@{}}{\textit{Path c}: Total Effect of Chatbot Condition $\rightarrow$ Goal Commitment} \\
\quad Constant & 4.14 & 0.05 & 85.73 & $< .001$ & [4.06, 4.22] \\
\quad Chatbot Condition & -0.06 & 0.02 & -2.95 & .003 & [-0.09, -0.03] \\
\quad Model $R^2$ & \multicolumn{5}{@{}l@{}}{.016, \textit{F}(1, 541) = 8.68, \textit{p} = .003} \\
\midrule
\multicolumn{6}{@{}l@{}}{\textbf{Mediation Effects (Bootstrap Estimates)}} \\
\midrule
\textbf{Effect} & \textbf{Estimate} & \textbf{Boot SE} & & & \textbf{90\% Boot CI} \\
\midrule
Total Effect (\textit{c}) & -0.06 & 0.02 & & .003 & [-0.09, -0.03] \\
Direct Effect (\textit{c'}) & -0.08 & 0.02 & & $< .001$ & [-0.11, -0.05] \\
Indirect Effect (\textit{ab}) & 0.02 & 0.01 & & & [0.01, 0.03] \\
\quad Completely Standardized & 0.05 & 0.01 & & & [0.02, 0.07] \\
\bottomrule

\multicolumn{6}{p{\linewidth}}{\small \textit{Note.} \textit{N} = 543. Coefficients represent effects per one-unit increase in chatbot sophistication (0 = ControlBot, 1 = GuidanceBot, 2 = SuggestionBot, 3 = FeedbackBot, 4 = GenBot). The significant indirect effect indicates that social presence partially mediates the relationship between chatbot sophistication and goal commitment, though the negative direct effect is larger in magnitude, resulting in a suppression pattern.} 
\end{tabular}
\Description{A linear mediation analysis table treating chatbot condition as an ordinal predictor of sophistication (0–4). Results show a suppression effect: increasing chatbot sophistication significantly increased Social Presence (Path a: B=0.16), which positively predicted Goal Commitment (Path b: B=0.14). However, the direct effect of sophistication on commitment was significantly negative (Path c': B=-0.08), meaning the small positive indirect effect (B=0.02) only partially mitigated an overall negative relationship.}
\end{table}

\FloatBarrier
\newpage
\section{\new{ Prompt Examples for the Different Conditions \textsuperscript{(2AC,R1)}}}\label{Control-bot-messages}
\begin{figure}[ht]
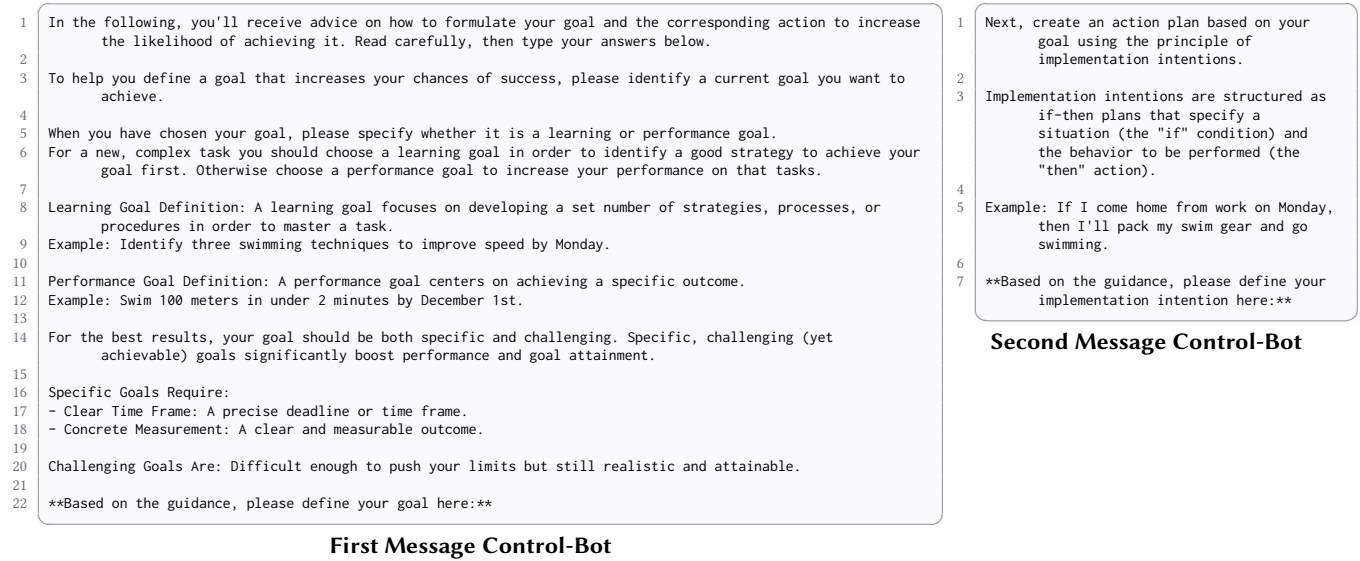

\centering

\begin{minipage}[t]{0.69\textwidth}
\begin{codecardGrey}[title=First Message Control-Bot]
In the following, you’ll receive advice on how to formulate your goal and the corresponding action to increase the likelihood of achieving it.  Read carefully, then type your answers below.

To help you define a goal that increases your chances of success, please identify a current goal you want to achieve.

When you have chosen your goal, please specify whether it is a learning or performance goal.
For  a new, complex task you should choose a learning goal in order to identify a good strategy to achieve your goal first. Otherwise choose a performance goal to increase your performance on that tasks.

Learning Goal Definition: A learning goal focuses on developing a set number of strategies, processes, or procedures in order to master a task.
Example: Identify three swimming techniques to improve speed by Monday.

Performance Goal Definition: A performance goal centers on achieving a specific outcome.
Example: Swim 100 meters in under 2 minutes by December 1st.

For the best results, your goal should be both specific and challenging. Specific, challenging (yet achievable) goals significantly boost performance and goal attainment.

Specific Goals Require:
- Clear Time Frame: A precise deadline or time frame.
- Concrete Measurement: A clear and measurable outcome.

Challenging Goals Are: Difficult enough to push your limits but still realistic and attainable.

**Based on the guidance, please define your goal here:**
\end{codecardGrey}
\end{minipage}\hfill
\begin{minipage}[t]{0.30\textwidth}
\begin{codecardGrey}[title=Second Message Control-Bot,]
Next, create an action plan based on your goal using the principle of implementation intentions. 

Implementation intentions are structured as if-then plans that specify a situation (the "if" condition) and the behavior to be performed (the "then" action).

Example: If I come home from work on Monday, then I’ll pack my swim gear and go swimming.

**Based on the guidance, please define your implementation intention here:**
\end{codecardGrey}
\end{minipage}

\caption{\new{\textbf{The two messages of Control Bot that are sent to the user.} These are hard-coded, but delivered through the same chat interface as all other conditions. The first message on the left explains goal setting theory and what makes a good goal. The second message on the right then instructs the user to set an implementation intention for the goal.}}
\Description{}
\label{fig:prompts_extended}
\Description{Two panels displaying the exact instructional scripts used for the Control Bot. The left panel ('First Message') defines Learning vs. Performance goals and outlines criteria for specific, challenging goals (Clear Time Frame, Concrete Measurement). The right panel ('Second Message') explains Implementation Intentions using an 'if-then' structure with a swimming example.}
\end{figure}

\begin{figure}[ht]
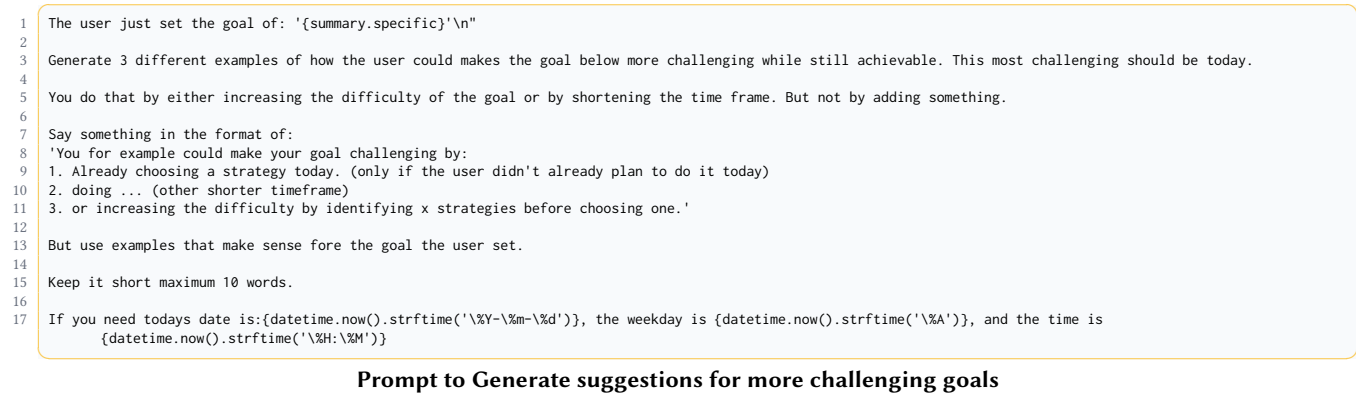

\centering
\begin{codecardYellow}[title=Prompt to Generate suggestions for more challenging goals]
The user just set the goal of: '{summary.specific}'\n"

Generate 3 different examples of how the user could makes the goal below more challenging while still achievable. This most challenging should be today.

You do that by either increasing the difficulty of the goal or by shortening the time frame. But not by adding something.

Say something in the format of:
'You for example could make your goal challenging by:
1. Already choosing a strategy today. (only if the user didn't already plan to do it today)
2. doing ... (other shorter timeframe)
3. or increasing the difficulty by identifying x strategies before choosing one.'

But use examples that make sense fore the goal the user set.

Keep it short maximum 10 words.

If you need todays date is:{datetime.now().strftime('\%Y-\%m-\%d')}, the weekday is {datetime.now().strftime('\%A')}, and the time is {datetime.now().strftime('\%H:\%M')}
\end{codecardYellow}
\caption{\new{\textbf{System prompt used to create more challenging performance goals} The model receives the user's previously set goal as context, along with the current date, and generates three potential options for making the goal more challenging. The resulting message is then appended to the static text used by the Suggestion Bot.}}
\Description{}
\label{fig:prompt_function_appendix}
\Description{A figure showing the system prompt text used to generate suggestions for making goals more challenging. The prompt instructs the AI to take the user's defined goal and output three variations that increase difficulty or shorten the timeframe (e.g., doing it "today"). It explicitly restricts the output to a maximum of 10 words per suggestion and includes the current date and time as context variables.}
\end{figure}

\FloatBarrier
\clearpage
\section{\new{Detailed participant demographics \textsuperscript{(R2)}}}
\label{sec:demographics}

\begin{table*}[h]
\centering
\caption{Participant Demographic Characteristics by Condition ($N = 543$)}
\label{tab:demographics}
\renewcommand{\arraystretch}{1.15}
\small

\begin{tabular}{@{}p{2.6cm}p{4.7cm}*{6}{c}@{}}
\toprule
\textbf{Demographic Variable} & \textbf{Category} &
\shortstack{\textbf{Overall}\\(N=543)} &
\shortstack{\textbf{ControlBot}\\(n=111)} &
\shortstack{\textbf{FeedbackBot}\\(n=108)} &
\shortstack{\textbf{SuggestionBot}\\(n=108)} &
\shortstack{\textbf{GuidanceBot}\\(n=109)} &
\shortstack{\textbf{GenBot}\\(n=107)} \\
\midrule

\multirow{2}{*}{\parbox[t]{2.6cm}{\textbf{Gender}}}
 & Male   & 268 (49.4) & 53 (47.7) & 55 (50.9) & 47 (43.5) & 55 (50.5) & 58 (54.2) \\
 & Female & 271 (49.9) & 57 (51.4) & 52 (48.1) & 60 (55.6) & 53 (48.6) & 49 (45.8) \\
\midrule

\multirow{6}{*}{\parbox[t]{2.6cm}{\textbf{Ethnicity}}}
 & White & 354 (65.2) & 70 (63.1) & 65 (60.2) & 76 (70.4) & 75 (68.8) & 68 (63.6) \\
 & Black or Black British & 74 (13.6) & 15 (13.5) & 17 (15.7) & 14 (13.0) & 13 (11.9) & 15 (14.0) \\
 & Asian or Asian British & 71 (13.1) & 19 (17.1) & 15 (13.9) & 11 (10.2) & 10 (9.2) & 16 (15.0) \\
 & Mixed or Multiple Ethnic Groups & 28 (5.2) & 4 (3.6) & 8 (7.4) & 1 (0.9) & 7 (6.4) & 8 (7.5) \\
 & Other Ethnic Group & 9 (1.7) & 1 (0.9) & 1 (0.9) & 4 (3.7) & 3 (2.8) & 0 (0.0) \\
 & Prefer not to say & 2 (0.4) & 0 (0.0) & 1 (0.9) & 1 (0.9) & 0 (0.0) & 0 (0.0) \\
\midrule

\multirow{5}{*}{\parbox[t]{2.6cm}{\textbf{Employment Status}\\{\footnotesize (available data: $n=451$)}}}
 & Full-time employed & 292 (53.8) & 54 (48.6) & 62 (57.4) & 57 (52.8) & 57 (52.3) & 62 (57.9) \\
 & Part-time employed & 72 (13.3) & 14 (12.6) & 13 (12.0) & 17 (15.7) & 11 (10.1) & 17 (15.9) \\
 & Not in paid work (e.g., homemaker, retired) & 33 (6.1) & 5 (4.5) & 5 (4.6) & 12 (11.1) & 5 (4.6) & 6 (5.6) \\
 & Unemployed and seeking work & 29 (5.3) & 7 (6.3) & 5 (4.6) & 5 (4.6) & 6 (5.5) & 6 (5.6) \\
 & Other (incl. due to start new job within one month) & 25 (4.6) & 3 (2.7) & 6 (5.6) & 3 (2.8) & 10 (9.2) & 3 (2.8) \\
\midrule

\multirow{2}{*}{\parbox[t]{2.6cm}{\textbf{Student Status}\\{\footnotesize (available data: $n=486$)}}}
 & Currently a student & 106 (21.8) & 19 (20.9) & 26 (26.5) & 16 (16.3) & 25 (25.3) & 20 (20.0) \\
 & Not a student & 380 (78.2) & 72 (79.1) & 72 (73.5) & 82 (83.7) & 74 (74.7) & 80 (80.0) \\
\bottomrule

\multicolumn{8}{p{\linewidth}}{\footnotesize \textit{Note.} Cells show $n$ (\%). For Gender and Ethnicity, percentages are calculated from the total participants in each condition. Student Status percentages are calculated from available Prolific responses in each condition.}
\end{tabular}
\Description{A demographic table for the 543 participants. The sample was gender-balanced (49.9\% Female, 49.4\% Male) and predominantly White (65.2\%), followed by Black (13.6\%) and Asian (13.1\%) participants. Regarding employment, the majority were full-time employed (64.7\%) and non-students (78.2\%). The mean age was 34.35 years (SD = 11.22).}
\end{table*}

\FloatBarrier
\section{\new{Overview of Measures\textsuperscript{(R1)}}}\label{appendix:measures}

{
\color{black}
\small
\renewcommand{\arraystretch}{1.4}

\begin{longtable}{
    >{\RaggedRight\hspace{0pt}}p{0.18\textwidth}
    >{\RaggedRight\hspace{0pt}}p{0.22\textwidth}
    >{\RaggedRight\hspace{0pt}}p{0.34\textwidth}
    >{\RaggedRight\hspace{0pt}\arraybackslash}p{0.20\textwidth}
}
\caption{\textcolor{black}{Overview of constructs, sources, item counts, and response formats. To avoid reproducing potentially copyrighted questionnaire materials, we report brief content summaries rather than full item wording for published multi-item scales. Self-constructed items (e.g., manipulation and attention checks) are shown verbatim. (R) indicates reverse-coded items.}} \label{tab:questionnaire_overview} \\
\toprule
\textbf{Construct} & \textbf{Scale source} & \textbf{What the items capture} & \textbf{Response format} \\
\midrule
\endfirsthead

\multicolumn{4}{c}{{\bfseries \textcolor{black}{\tablename\ \thetable{} -- continued}}} \\
\toprule
\textbf{Construct} & \textbf{Scale source} & \textbf{What the items capture} & \textbf{Response format} \\
\midrule
\endhead

\midrule
\multicolumn{4}{r}{{\footnotesize \textit{Continued on next page}}} \\
\bottomrule
\endfoot

\bottomrule
\endlastfoot

\textbf{Goal Commitment}
& Adapted from \citet{hollenbeck_empirical_1989}
& 5 items (2R). Commitment strength, importance of attainment, and willingness to persist or not abandon the goal.
& 5-point Likert (Strongly disagree -- Strongly agree) \\
\midrule

\textbf{Perceived Goal Difficulty} \newline \textit{\scriptsize (vs.\ others with similar skills)}
& 4 items from \citet{lee_exploring_1992}
& 4 semantic-differential dimensions: perceived challenge, required effort, required thought/skill, and required persistence.
& 5-step semantic differential (anchors per dimension) \\
\midrule

\textbf{Perceived Social Presence} \newline \textit{\scriptsize (regarding the chatbot)}
& Adapted from~\citet{kumar_research_2006}
& 4 items. Perceived human-likeness and interpersonal presence (e.g., felt personal, sociable, and sensitive).
& 7-point Likert (Strongly disagree -- Strongly agree) \\
\midrule

\textbf{Manipulation Check}
& Self-constructed 
& 4 items (incl.\ attention check), shown verbatim: \newline
(1) Examples were adapted to my goal context. \newline
(2) Advice was broken into small, easy steps. \newline
(3) Chatbot answered questions and flagged missing details. \newline
(4) \textit{Attention check:} Please select ``Disagree'' for this question.
& 5-point Likert (Strongly disagree -- Strongly agree) \\
\midrule

\textbf{Performance Expectancy}
& UTAUT2: \citet{venkatesh_consumer_2012}
& 5 items. Expected usefulness and effectiveness of the chatbot for goal achievement.
& 5-point Likert (Strongly disagree -- Strongly agree) \\
\midrule

\textbf{Effort Expectancy}
& UTAUT2: \citet{venkatesh_consumer_2012}
& 6 items (1R). Perceived ease of use, clarity, and time/effort required.
& 5-point Likert (Strongly disagree -- Strongly agree) \\
\midrule

\textbf{Hedonic Motivation}
& UTAUT2: \citet{venkatesh_consumer_2012}
& 3 items. Enjoyment and fun while using the chatbot.
& 5-point Likert (Strongly disagree -- Strongly agree) \\
\midrule

\textbf{Attitude toward Chatbot}
& UTAUT2: \citet{venkatesh_consumer_2012}
& 5 items (1R). Overall positive or negative evaluation of using the chatbot.
& 5-point Likert (Strongly disagree -- Strongly agree) \\
\midrule

\textbf{Goal Attainability}
& Self-constructed 
& 1 item. Perceived attainability of the goal just set.
& Single item (scale anchors as in survey) \\
\midrule

\textbf{Intention to Act}
& Adapted from \citet{fishbein_predicting_2011}
& 1 item. Intention to carry out the defined action.
& Slider (Unlikely -- Likely) \\
\midrule

\textbf{Goal Specificity} \newline \textit{\scriptsize (expert-coded)}
& Coding scheme 
& Timeframe (0--2) + measurability (0--2), summed to 0--4.
& Two coders; ordinal scores \\
\midrule

\textbf{Implementation-Intention Quality} \newline \textit{\scriptsize (expert-coded)}
& Coding scheme 
& Cue present, action present, explicit cue--action link, time/place specified (0--4 total).
& Two coders; ordinal scores \\
\Description{A comprehensive table listing the specific survey items and response scales for nine constructs: Goal Commitment, Perceived Goal Difficulty, Perceived Social Presence, Manipulation Check, Performance Expectancy, Effort Expectancy, Hedonic Motivation, Attitude towards Chatbot, and Intention to Act. Most constructs utilize a 5-Point Likert scale, while Social Presence uses a 7-Point scale, Goal Difficulty uses a Semantic Differential scale, and Intention to Act uses a probability slider.}
\end{longtable}
}

\end{document}